\documentclass[fleqn,usenatbib]{mnras}

\usepackage{newtxtext,newtxmath}
\usepackage[T1]{fontenc}

\DeclareRobustCommand{\VAN}[3]{#2}
\let\VANthebibliography\thebibliography
\def\thebibliography{\DeclareRobustCommand{\VAN}[3]{##3}\VANthebibliography}

\usepackage{graphicx}
\usepackage{amsmath}
\usepackage{amsfonts}
\usepackage{amsbsy}
\usepackage{enumitem}
\usepackage{physics}
\usepackage{siunitx}
\usepackage{afterpage}
\usepackage{stmaryrd}
\usepackage[x11names]{xcolor}
\usepackage{array}
\usepackage{hyphenat}
\usepackage[ruled,vlined]{algorithm2e}
\usepackage[noend]{algpseudocode}
\usepackage[makeroom]{cancel}

\DeclareMathOperator*{\argmin}{argmin}

\newcommand{\lnb}[1]{%
  \ln\left[#1\right]%
}

\newcommand{\carpool}{CWAV20}

\title[CARPool Covariance]{CARPool Covariance: Fast, unbiased covariance estimation for large-scale structure observables}

\author[Nicolas Chartier and Benjamin D. Wandelt]{
\parbox{\textwidth}{
\LARGE
Nicolas Chartier$^{1, 2}$
and Benjamin D.\ Wandelt $^{2,3}$
}
\vspace{0.4cm}
\\
$^{1}$Laboratoire de Physique de l'\'{E}cole Normale Sup\'{e}rieure, ENS, Universite PSL, CNRS, Sorbonne Universit\'{e}, Universit\'{e} de Paris, F-75005 Paris, France\\
$^{2}$Sorbonne Universit\'{e}, CNRS, UMR 7095, Institut d'Astrophysique de Paris, 98 bis bd Arago, 75014 Paris, France\\
$^{3}$Center for Computational Astrophysics, Flatiron Institute, 162 5th Avenue, New York, NY 10010, USA
}

\date{Accepted 2021 October 25. Received 2021 October 25; in original form 2021 October 18}
\pubyear{2021}

\begin{document}

\label{firstpage}
\pagerange{\pageref{firstpage}--\pageref{lastpage}}
     
\maketitle
\begin{abstract}
The covariance matrix $\boldsymbol{\Sigma}$ of non-linear clustering statistics that are measured in current and upcoming surveys is of fundamental interest for comparing cosmological theory and data and a crucial ingredient for the likelihood approximations underlying widely used parameter inference and forecasting methods.
The extreme number of simulations needed to estimate  $\boldsymbol{\Sigma}$ to sufficient accuracy poses a severe challenge. Approximating  $\boldsymbol{\Sigma}$ using inexpensive but biased \textit{surrogates} introduces model error with respect to full simulations, especially in the non-linear regime of structure growth. To address this problem we develop a matrix generalization of Convergence Acceleration by Regression and Pooling (CARPool)  to combine a small number of simulations with fast surrogates and obtain low-noise estimates of $\boldsymbol{\Sigma}$ that are unbiased by construction.
Our numerical examples use CARPool to combine \texttt{GADGET-III} $N$-body simulations with fast surrogates computed using COmoving Lagrangian Acceleration (COLA). Even at the challenging redshift $z=0.5$, we find variance reductions of at least $\mathcal{O}(10^1)$ and up to $\mathcal{O}(10^4)$ for the elements of the matter power spectrum covariance matrix on scales $8.9\times 10^{-3}<k_\mathrm{max} <1.0$ $h {\rm Mpc^{-1}}$. We demonstrate comparable performance for the covariance of the matter bispectrum, the matter correlation function and probability density function of the matter density field. We compare eigenvalues, likelihoods, and Fisher matrices computed using the CARPool covariance estimate with the standard sample covariance and generally find considerable improvement except in cases where $\boldsymbol{\Sigma}$ is severely ill-conditioned. 
\end{abstract}


\section{Introduction}
\label{sec:introduction} 
In the era of precision cosmology, modelling the statistical properties of observables is crucial to derive cosmological parameters constraints from large-scale structure surveys. Particularly, the covariance matrix $\boldsymbol{\Sigma}$ of clustering statistics, such as the matter power spectrum and the matter bispectrum, as well as its inverse---the precision matrix---are key elements when building likelihood approximations, efficient estimators, or developing optimal summaries of observations for cosmological inference (\cite{moped,refId0,2009ApJ...700..479T,2012MNRAS.426.1262H,2013PhRvD..88f3537D,2013MNRAS.431.3349H,2014MNRAS.442.2728T,2014MNRAS.439.2531P,10.1093/mnras/stu2190,alsingwandelt2018,2019A&A...631A.160H,2020PhRvD.102h3514H,2020PhRvD.102l3521W}).

Unfortunately, estimating the covariance matrix of large scale structure observables is  extremely challenging owing to both the large number of samples required and the computational cost per sample. 
A brute force solution to estimate covariance matrices would  be to generate mock samples of survey statistics with computationally intensive $N$-body simulations reproducing the conditions of observation (volume, redshifts, sky area...), and then to compute the sample covariance matrix, which is an unbiased and positive (semi-) definite estimator of the true covariance. But high quality estimates of the covariance are necessary because we actually need its inverse, the precision matrix. This will be dominated by the smallest eigenvalues of the covariance. It is just these small eigenvalues that need the largest number of samples to converge. For example, \citet{2016MNRAS.458.4462B} found in numerical experiments for a Euclid-like survey that at least 5,000 independent $N$-body simulations are needed to estimate the power spectrum covariance in order to obtain cosmological parameter forecasts at an adequate level of accuracy given the precision of upcoming surveys. 
In spite of recent progress in optimization of various $N$-body codes, with GPU-acceleration and distributed-memory solutions \citep{2005MNRAS.364.1105S,2009PASJ...61.1319I,2013arXiv1310.4502W,phdAbacus,2016NewA...42...49H,2017ComAC...4....2P}, limited CPU time and memory resources and the large number of samples required mean that it will remain impractical to rely solely on full $N$-body simulations for covariance matrix estimation. For this reason, cosmologists have been investigating less costly alternatives to tackle next-generation datasets.

For certain clustering statistics that are amenable to perturbative treatment, analytical predictions allow computing covariance estimates rapidly, at the cost of making assumptions on the survey data. Such predictions typically use the Gaussian limit for the covariance. For example, \citet{2020MNRAS.491.3290P} developed the \texttt{RascalC} code that estimates the covariance of the two-point galaxy correlation function (2PCF) and only needs one dataset as input: a shot noise rescaling, constrained by jackknife covariance matrices, describes deviations from Gaussianity, and as a result the large scale model covariance matrix is fully consistent with mocks. \citet{2019MNRAS.490.5931P} had applied a similar approach to the auto- and cross-covariances of 2PCFs and three-point correlation functions (3PCFs) for general real-space survey statistics. \citet{2019JCAP...01..016L} proposed an analytical computation of the "disconnected" (Gaussian) part of the covariance, that is dominant on large scales, for both correlation functions and their Fourier space counterpart (power spectra): they found valuable accordance with mock estimates. \citet{2017MNRAS.466..780M} also experimented with an analytical decomposition of the the covariance matrix motivated by perturbation theory, and managed to get a $10\%$ agreement with simulations up to $k \approx 1$ $h {\rm Mpc^{-1}}$.
\citet{Mohammed:2014lja} tested a simple model for the matter power spectrum motivated by  the Zel'dovich approximation, and stressed the influence of the simulation box volume on the convergence of the covariance matrix. Useful reviews of methods using theoretical predictions include \citet{Bernardeau:2001qr} and \citet{Desjacques:2016bnm}.

Alternatively, computational cosmologists have proposed various approximate solvers designed to be much faster than $N$-body simulations.
One class of such methods exploits the availability of low order Lagrangian Perturbation Theory (LPT): \citet{2002MNRAS.329..629S} (PTHalos), \citet{2012JCAP...04..013T}, \citet{2013MNRAS.433.2389M} inspired by \citet{pino} (PINOCCHIO), or \citet{2015MNRAS.446.2621C} (EZmocks).
Futhermore, a significant number of Particle-Mesh (PM) codes, which treat the force as a field on a mesh, use the large-scale approximation provided by LPT: \citet{Tassev_2013} (COLA), \citet{2015arXiv150207751T} (sCOLA) implemented by \citet{2020A&A...639A..91L}, \citet{2016MNRAS.463.2273F} (FastPM) available in a distributed version by \citet{2020arXiv201011847M}, \citet{2014MNRAS.437.2594W} (QPM), and \citet{2014MNRAS.439L..21K} (PATCHY), to name a few.

Another family of approaches comprises mathematical models with free-parameters--\textit{emulators}--that are trained on simulation suites covering a given range of cosmologies to then directly predict clustering parameters from upcoming data. Recent studies include \citet{DeRose_2019} \citet{McClintock_2019}, \citet{Zhai_2019}, \citet{2019arXiv190713167M}, \citet{2020arXiv200108055K} or \citet{2020arXiv200406245A}. 
Although they can provide lightning-fast estimates, emulators are limited by the parameter range of the training set and do not guarantee unbiased results with respect to full solvers, of which they still need a large number of realisations to train. Some emulators based on deep learning architectures have been shown to reproduce particle positions or matter density fields from input initial conditions. In other words, they can produce snapshots of a low-resolution cosmological $N$-body code from which any clustering statistics can be extracted: \citet{2020arXiv201002926D}, \citet{He_2019}, \citet{Kodi_Ramanah_2020}.
Fast approximate solvers -- which we will refer to as \textit{surrogates} and which comprise all the previous families of methods we mentioned -- unfortunately do not match the accuracy of full $N$-body mocks, especially in the deeply non-linear regime. \citet{2019MNRAS.482.1786L}, \citet{2019MNRAS.485.2806B} and \citet{2019MNRAS.482.4883C} find statistical biases in parameters estimation with covariance matrices from surrogates up to $20\%$ higher than with covariances from full $N$-body codes.

Some works in cosmology are specifically dedicated to improving the estimation of covariance matrices. In the particular case of Gaussian-distributed weak lensing power spectra, \citet{2013MNRAS.432.1928T} assessed the limits on parameter estimation imposed by the accuracy of the precision matrix, and discussed solutions to relax them. \citet{10.1093/mnras/stv2259} implemented a technique called \textit{tapering} to estimate covariance and precision matrices and proved to be successful in reducing the confidence intervals of parameters without introducing bias. \citet{2016MNRAS.457..993P} fitted a theoretically-motivated model with a mock catalogue in order to estimate the covariance matrix with fewer samples. \citet{2020arXiv200413436F} provided more insight on the impact of jackknife resampling on covariance matrix estimates, which had also been experimented with by \citet{2016arXiv160600233E} for the two-point galaxy clustering correlation function. In \citet{2019MNRAS.483..189H}, using a likelihood conditioned on both theoretical and simulated covariance matrices of summary statistics reduced the required number of simulations for covariance estimation.
\citet{2008MNRAS.389..766P} applied the concept of linear shrinkage to the matter power spectrum covariance matrix: by optimally combining an empirical estimate with a specified simple target (for instance a diagonal covariance), they significantly improved the estimated covariance when few simulations are available. Regarding non-linear shrinkage, see for instance \citet{2017MNRAS.466L..83J}.
In addition, the precision matrix being essential to derive parameters confidence bounds, the fact that the inverse of the unbiased covariance estimator is not an unbiased estimator of the precision is now well known notably thanks to \citet{2007A&A...464..399H}. Numerous cosmology studies focus on precision matrix estimation and on the effects -- parameters shifts for instance -- of precision matrix biases (\citet{2018MNRAS.473.4150F},  \citet{2018MNRAS.473.2355S} \citet{2021PhRvD.103d3508P}, \citet{2021arXiv210810402P}, \citet{2021MNRAS.tmp.2208F}).

\citet{2020arXiv200908970C} (\carpool\ from now on) developed the Convergence Acceleration by Regression and Pooling (CARPool) method, a general approach to reducing the number of simulations needed for low variance and explicitly unbiased estimates of clustering statistics. Equivalently, CARPool can be viewed as a way to obtain unbiased results from fast surrogates by running a small number of simulations. \carpool\ demonstrated a dramatic reduction of the number of simulations required to estimate the mean of a given statistic by exploiting the variance reduction principle known as \textit{control variates} and combining a smaller number of costly simulations with a larger number of \textit {surrogates}. CARPool exploits the correlation between full $N$-body runs and fast surrogates run on the same initial conditions, and proved to be very efficient for the estimation of the mean of the matter power spectrum, the bispectrum or the one-point probability density function (PDF).

It is therefore natural to study whether CARPool can improve the estimation of covariance matrices while reducing the number of simulations\footnote{ \citet{2016PhRvD..93j3519P}, \citet{2016MNRAS.462L...1A}, and \citet{Villaescusa_Navarro_2018} discuss variance reduction by designing special initial conditions that explicitly bias certain higher-order n-point functions low. These would therefore not seem promising for improving estimates of the covariance matrix.}. Showing this to be the case is the main contribution of this paper.   We will first recall some theoretical results and generalize the CARPool approach to covariance matrices in section \ref{sec:methods}. Then, we will show in section \ref{sec:experiments}, using  likelihoods, eigenvalues, and Fisher information matrices computed from the estimated covariance matrices, that CARPool can reduce the number of mocks needed to estimate covariance matrices of clustering statistics by at least one order of magnitude and, depending on scale and observable, in some cases by several orders of magnitude. We will discuss our results and conclude in  section \ref{sec:conclusions}.

\section{Statistical methods}\label{sec:methods}

We adopt the same notation system as in \carpool\ up to  a few necessary adaptations. Namely, let $\boldsymbol{y}$ be a costly simulation observable, constituted by scalar measurements $y_i, 1 \leq i \leq p$, such that $\mathbb{E} \left[ \boldsymbol{y} \right] = \boldsymbol{\mu} \in \mathbb{R}^p$, and  $\boldsymbol{c}$ an approximate random observable with $\mathbb{E} \left[ \boldsymbol{c} \right] = \boldsymbol{\mu_c} \in \mathbb{R}^{q}$. In computational cosmology, $\boldsymbol{y}$ and $\boldsymbol{c}$ can be, for instance, Cold Dark Matter (CDM) power spectra with $p$ and $q$ power band bins, respectively.

\subsection{Variance reduction with control variates}
In this section, we recall some results dealing with the variance reduction technique known as \textit{control variates}. For more details, see \carpool.

\subsubsection{General multivariate case}\label{sec:general}
In order to compute an unbiased estimator $\boldsymbol{\hat{\mu}}$ of $\boldsymbol{\mu}$, a straightforward solution is to use the sample mean from a set of \textit{independent and identically distributed} realisations $\boldsymbol{y}_n$, $n=1,\dots N$,

\begin{equation}\label{eq:sampMean}
\boldsymbol{\hat{\mu}} = \boldsymbol{\bar{y}} \equiv \frac{1}{N}\sum_{n=1}^{N} \boldsymbol{y}_n\,.
\end{equation}

The standard deviation $\sigma_{i}$ of each $\bar{y}_i,  1\leq i\leq p$,  decreases slowly as $\mathcal{O}(N^{-\frac{1}{2}})$ when the number $N$ of available samples increases. We are interested in computing a more precise and unbiased estimator of $\boldsymbol{\mu}$ so that we need less simulations $\boldsymbol{y}_n$ and thus less computational resources.

To this end, we can use fast surrogates  $\boldsymbol{c}_n$ that are correlated with the costly simulations $\boldsymbol{y}_n, n=1,\dots N$ by constructing the random vectors
\begin{equation}\label{eq:mvCV}
    \boldsymbol{x}_n(\boldsymbol{\beta}) = \boldsymbol{y}_n - \boldsymbol{\beta} \left( \boldsymbol{c}_n - \boldsymbol{\mu_c} \right)\,,
\end{equation}
with \textit{control matrix} $\boldsymbol{\beta} \in \mathbb{R}^{p \times q}$.

The \textit{control variates} estimator is then the sample mean of $N$ samples from equation \eqref{eq:mvCV}:
\begin{equation}\label{eq:sampleCV}
    \boldsymbol{\hat{\mu}}(\boldsymbol{\beta}) = \boldsymbol{\bar{x}}(\boldsymbol{\beta}) = \boldsymbol{\bar{y}} - \boldsymbol{\beta} \left( \boldsymbol{\bar{c}} - \boldsymbol{\mu_c} \right)\,.
\end{equation}
This estimator is unbiased by construction, $\mathbb{E}\left[\boldsymbol{\hat{\mu}}(\boldsymbol{\beta})\right] = \boldsymbol{\mu}$ regardless of any bias in the surrogates. The unique choice 
\begin{equation}\label{eq:betaStarMV}
    \boldsymbol{\beta^{\star}} = \argmin_{\boldsymbol{\beta} \in \mathbb{R}^{p \times q}} \det \left(\boldsymbol{\Sigma_{xx}}(\boldsymbol{\beta})\right) = \boldsymbol{\Sigma_{yc}}\boldsymbol{\Sigma_{cc}^{-1}}\,.
\end{equation}
gives the minimum variance estimator (\citet{10.1287/opre.33.3.661}; see \carpool\ for a Bayesian derivation).

As shown in \carpool, for many practical applications where $p>>1$, imposing structure on $\boldsymbol{\beta}$ can lead to large reductions in computational cost when $\boldsymbol{\beta}$ must be estimated from simulation/surrogate pairs. In the following, we will first consider diagonal $\boldsymbol{\beta}$ and then the more general case of sparse $\boldsymbol{\beta}$.

\subsubsection{Diagonal case}\label{sec:uni}
When $\boldsymbol{\beta}$ is diagonal, the problem reduces to estimating $p$ independent quantities. It is easy to prove (\carpool) that there exists a single control coefficient $\beta^{\star}$ that minimizes the variance of the resulting random variable
\begin{equation}
\begin{aligned}\label{eq:uniBeta}
     x(\beta) &= y - \beta \left( c - \mu_c \right)\\
    \beta^{\star} &=\argmin_{\beta \in \mathbb{R}} \sigma_{x(\beta)}^2= \frac{\mathrm{cov}(y,c)}{\sigma_c^2}\,,
\end{aligned}
\end{equation}
and with subsequent variance reduction
\begin{equation}
    \frac{\sigma_{x(\beta^{\star})}^2}{\sigma_y^2} = 1 - \rho_{y,c}^2\,.
\end{equation}
$\rho_{y,c}$ is the Pearson correlation coefficient between $y$ and $c$. 
This  case is equivalent to estimating the multivariate quantity $\boldsymbol{\mu}$, equation \eqref{eq:sampleCV}, while replacing $\boldsymbol{\beta^{\star}}$ with

\begin{equation}\label{eq:betaDiag}
\boldsymbol{\beta^{diag}} = \mathrm{diag}\left(
\frac{cov(y_1,c_1)}{\sigma_{c_{1}}^2},\frac{cov(y_2,c_2)}{\sigma_{c_{2}}^2},\dots ,\frac{cov(y_p,c_p)}{\sigma_{c_{p}}^2}\right)\,.
\end{equation}

\carpool\ applied this diagonal case to the estimation of the mean of the matter power spectrum, the matter bispectrum and the one-dimensional matter probability density function (PDF).

\subsubsection{Sparse case}\label{sec:hb}
In some applications, multiple elements in $\boldsymbol{c}$ can help reduce variance of any element of $\boldsymbol{y}$. In this case the problem reduces to  $p$ separate estimates with $q$ control variates each,
\begin{equation}\label{eq:hybridCV}
    x(\beta) = y - \boldsymbol{\beta^T} \left( \boldsymbol{c} - \boldsymbol{\mu_c} \right).
\end{equation}
 with $\boldsymbol{\beta} \in \mathbb{R}^q$.
The optimal choice is 
\begin{equation}\label{eq:hybridBeta}
     \boldsymbol{\beta_m^{\star}} = \boldsymbol{\Sigma_{cc}^{-1}}\boldsymbol{\sigma}_{y\boldsymbol{c}}
\end{equation}
with $\boldsymbol{\sigma}_{y\boldsymbol{c}}$ the $q$-dimensional column vector of covariances defined by $\boldsymbol{\sigma}_{y\boldsymbol{c}}[i] = cov(y,c_i), 1 \leq i \leq q$ \citep{Lavenberg1981APO, 10.1287/opre.33.3.661,10.1007/978-3-642-56046-0_3}. 

\begin{figure}
    \includegraphics[width=\columnwidth]{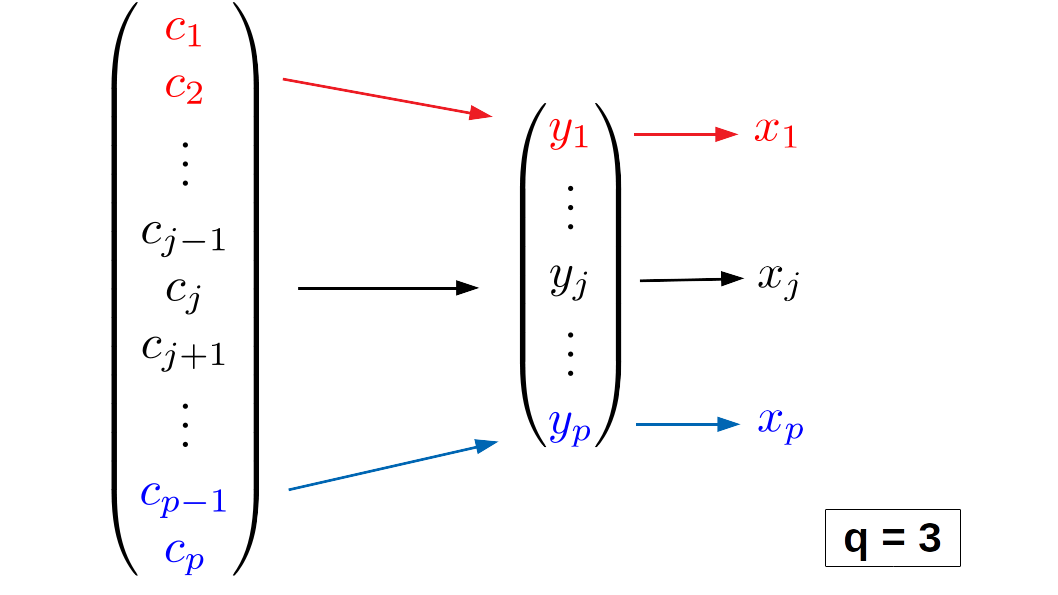}
    \caption{Illustration of sparse control matrix $\boldsymbol{\beta}$ using $q=3$  elements of the surrogate $\boldsymbol{c}$ to reduce variance on an element of  the simulation $\boldsymbol{y}$. In this example we use adjacent elements of the surrogate; the boundary cases have $q=2$.}\label{fig:multiple}
\end{figure}

The attainable variance reduction  with $\boldsymbol{\beta_m^{\star}}$ is
\begin{equation}\label{eq:varHbCV}
    \frac{\sigma_{x(\boldsymbol{\beta_m^{\star}})}^2}{\sigma_y^2} = 1 - \frac{\boldsymbol{\sigma}_{y\boldsymbol{c}}^{\boldsymbol{T}}\boldsymbol{\Sigma_{cc}^{-1}}\boldsymbol{\sigma}_{y\boldsymbol{c}}}{\sigma_y^2}\,.
\end{equation}
The Bayesian derivation in \carpool\  explains the form of the right hand side of equation \eqref{eq:varHbCV} as the ratio of the conditional and marginal covariances of $y$. The conclusion remains the same as before: the more correlation, the smaller the variance.  
The idea is to use more of the elements of $\boldsymbol{c}$ to improve the variance reduction on $y$ at the cost of estimating the vector $\boldsymbol{\beta_m^{\star}}$ from equation \eqref{eq:hybridBeta} instead of the scalar $\beta$ from equation \eqref{eq:uniBeta}. Figure \ref{fig:multiple} illustrates the principle for $q=3$ neighbouring surrogate variables per simulation variable. From now on, we will consider $\mathrm{dim}(\boldsymbol{y}) = \mathrm{dim}(\boldsymbol{c}) = p$ and have $q$ designate the number of surrogates elements taken for each $y$ in the sparse case.

\subsection{Application to covariance}
\label{sec:appcov}
In this section we rewrite the covariance estimation problem as an instance of the CARPool method described above. We can replace the vectors $\boldsymbol{y}$ and $\boldsymbol{c}$ of size $p$ by $\boldsymbol{Y}$ and $\boldsymbol{C}$ of size $P=p(p+1)/2$, representing the covariance matrix elements. If $\left\{\boldsymbol{y}_1,\dots,\boldsymbol{y}_N\right\}$ are realisations of the random vector $\boldsymbol{y}$, then the samples $\left\{\boldsymbol{y}_1-\boldsymbol{\bar{y}}, \dots, \boldsymbol{y}_N-\boldsymbol{\bar{y}}\right\} \equiv \left\{ \boldsymbol{\widetilde{y}}_1, \dots, \boldsymbol{\widetilde{y}}_N\right\}$ are also realisations of a multivariate random variable; and so are the outer products $\left\{ \boldsymbol{\widetilde{y}}_1 \otimes \boldsymbol{\widetilde{y}}_1, \dots,  \boldsymbol{\widetilde{y}}_N \otimes \boldsymbol{\widetilde{y}}_N \right\} \equiv \left\{\boldsymbol{Y}_1, \dots, \boldsymbol{Y}_N \right\}$ with $P$ unique elements.
We rewrite the sample covariance matrix in terms of these outer products
\begin{equation}\label{eq:sampCov}
    \begin{aligned}
    \boldsymbol{\widehat{\Sigma}_{yy}} &= \frac{1}{N-1} \sum_{i=1}^N \left( \boldsymbol{y}_i -\boldsymbol{\bar{y}} \right) \left( \boldsymbol{y}_i -\boldsymbol{\bar{y}} \right)^{\boldsymbol{T}}\\ &\equiv \frac{N}{N-1} \times \frac{1}{N}\sum_{i=1}^N \boldsymbol{Y}_i\\ &\equiv \gamma \boldsymbol{\overline{Y}} \,,
    \end{aligned}
\end{equation}
with Bessel's correction factor $\gamma=N/(N-1)$. This is an unbiased estimator of the true covariance $\boldsymbol{\Sigma_{yy}}$.
Similarly,  the surrogate samples $\boldsymbol{C}_i$ have sample mean
\begin{equation}\label{eq:sampSurr}
    \boldsymbol{\overline{C}}  \equiv \frac{1}{\gamma} \boldsymbol{\widehat{\Sigma}_{cc}} =\frac{1}{N}\sum_{i=1}^N \boldsymbol{C}_i  \,.
\end{equation}
Note the additional constraint of $\boldsymbol{y}$ and $\boldsymbol{c}$ having finite fourth-order moments (i.e variance of the covariance). 

\begin{figure}
    \includegraphics[width=\columnwidth]{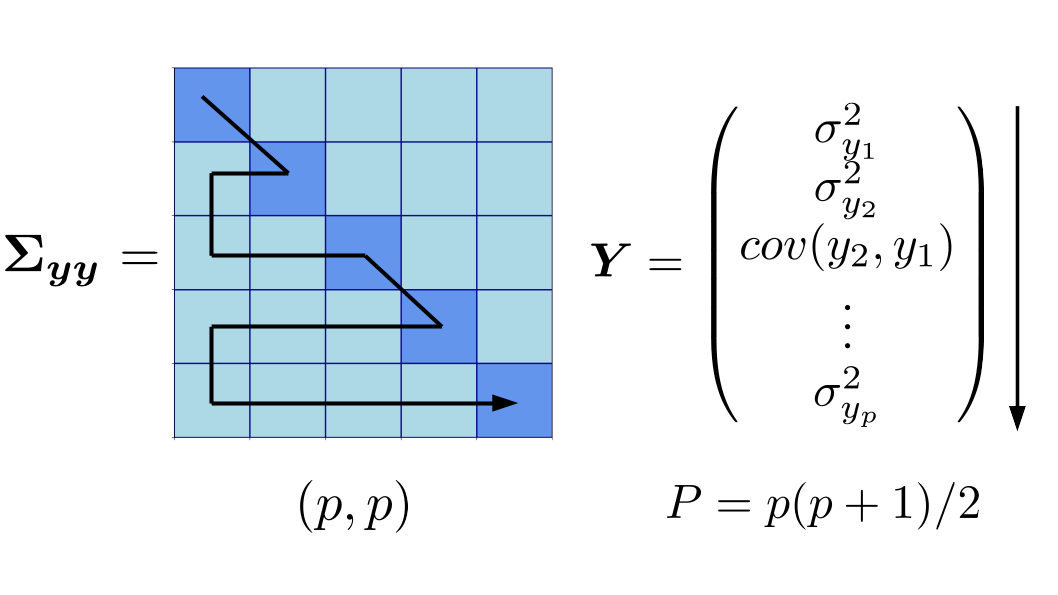}
    \caption{We generalize the CARPool method to covariance matrix estimation using the symmetric matrix vectorization above. This turns a  $p\times p$ symmetric matrix into a vector of its $p(p+1)/2$ unique elements.}
    \label{fig:vectorization}
\end{figure}

Figure \ref{fig:vectorization} explains how we build such vectors from symmetric matrices by ensuring matrix elements remain neighbors in the resulting vector\footnote{Alternatively, \texttt{numpy} provides the function \texttt{tril$\_$indices} (\texttt{triu$\_$indices}) to simply extract the lower (upper) triangular part of a $2D$-array in a row-major order}.\\
For simplicity, in equations \eqref{eq:sampCov}, \eqref{eq:sampSurr} and \eqref{eq:sampCARPcov}, we have identified the capital bold vectors of size $P$ with their reconstruction into a symmetric $(p,p)$ matrix to emphasize that the CARPool covariance matrix can be framed as a standard CARPool estimate. From this point, we also drop the "hat" of the estimated quantities for notational simplicity. 

The CARPool covariance estimate is then simply an instance of the CARPool method, equation  \eqref{eq:sampleCV}, applied to the vectorized sample covariances
\begin{equation} \label{eq:sampCARPcov}
\begin{aligned}
    \boldsymbol{\Sigma_{yy}}(\boldsymbol{\beta}) =\gamma \boldsymbol{\overline{X}}(\boldsymbol{\beta}) = \gamma  \boldsymbol{\overline{Y}} &- \gamma \boldsymbol{\beta}\left(\boldsymbol{\overline{C} - \boldsymbol{\mathcal{M}}_{C}}\right)\,,
    \end{aligned}
\end{equation}
where $\boldsymbol{\mathcal{M}}_{C}$ plays the role of $\boldsymbol{\mu_c}$ in the case of covariance estimation, that is to say it is the surrogate covariance matrix computed from a separate set of surrogate realisations.
The estimator is unbiased by construction, $\mathbb{E}\left[ \boldsymbol{\Sigma_{yy}}(\boldsymbol{\beta}) \right] = \boldsymbol{\Sigma_{yy}}$,
and corrected by the factor $\gamma$.  

Note that the sample covariance and the unbiased estimator of the covariance from equation \eqref{eq:sampCARPcov} do not necessarily yield, when inverted, an unbiased estimate of the precision. \citet{2007A&A...464..399H} point out a correction factor (known as the "Hartlap factor" in cosmology) for data sampled from a multi-variate Gaussian distribution that has been widely used. In recent works, impacts of biases in the estimated precision matrix up until parameter constraints and shifts, as well as methods to improve the estimation, have been emphasized, e.g., in \citet{2018MNRAS.473.4150F}, \citet{2021arXiv210810402P} and \citet{2021PhRvD.103d3508P}. In the following numerical experiments, we will assess the performance of the inverse of our newly proposed covariance estimate through multiple tests: by visual comparison, through studying the eigenvalues of the covariances  (which are simply the reciprocals of the eigenvalues of the precision) and by computing the Fisher matrices for parameter constraints, which are computed through the precision. A detailed description of tests will be given in section \ref{sec:testsDescription}.

Also, unlike the sample covariance matrix, the estimate $\boldsymbol{\Sigma_{yy}}(\boldsymbol{\beta})$ is not guaranteed to be positive semi-definite by construction for a finite number of samples, even though its expectation is. We will comment further on positive-definiteness in the numerical results of section \ref{sec:results}.

\section{Numerical experiments}\label{sec:experiments}

Our numerical analysis assume a $\Lambda$ Cold Dark Matter ($\Lambda$CDM) cosmology  congruent with the {\it Planck} constraints from \citet{2020A&A...641A...6P}: $\Omega_\mathrm{m}=0.3175$, $\Omega_\mathrm{b}=0.049$, $h=0.6711$, $n_s=0.9624$, $\sigma_8=0.834$, $w=-1.0$ and $M_{\nu}=0.0~\mathrm{eV}$.
For a reminder about how to apply CARPool, see Figure 1 in \carpool: the principle stays the same, except the vectorized outer products of centered data, as explained above, play the role of the data samples to estimate the covariance matrix.
The numerical analysis presented below compares the following unbiased covariance matrix estimators:
\begin{itemize}
    \item \texttt{GADGET}, where we compute the sample covariance $\boldsymbol{\Sigma_{yy}}$ with equation \eqref{eq:sampCov} from $N$-body simulations only.
    \item Diagonal CARPool applied individually to each unique covariance matrix element estimator, with equation \eqref{eq:uniBeta} applied to the vectorized covariance $\boldsymbol{X}$. This framework can simply be referred to as "$q=1$".
    \item Sparse CARPool, where we estimate each simulation covariance matrix element with $q>1$ surrogate matrix elements according to equation \eqref{eq:hybridCV}.
\end{itemize}
We stress that for the CARPool estimate, we compute the control matrix  $\boldsymbol{\beta}$ from  the same $N$ simulations entering $\boldsymbol{\bar{Y}}$ in equation \eqref{eq:sampCARPcov}.

\subsection{Simulation and surrogate data}
We briefly describe here the chosen simulation and surrogate solvers we use; for details please refer to the numerical experiments of \carpool\ where the same simulations and surrogates are used.
The solvers evolve $\mathcal{N}_\mathrm{p} = 512^3$ CDM particles in a box volume of $\left( 1000~h^{-1} {\rm Mpc} \right)^3$.
The simulation-surrogate sample pairs take the same Second-order Lagrangian perturbation theory (2LPT) initial conditions at starting redshift $z_{s}=127.0$.

\subsubsection{$N$-body solver}
The simulation outputs are part of the publicly available \textit{Quijote} simulation suite\footnote{ \url{https://github.com/franciscovillaescusa/Quijote-simulations}} \citep{2020ApJS..250....2V}. The full $N$-body simulations were run with the TreePM code \texttt{GADGET-III}, stemming from \texttt{GADGET-II} by \citet{2005MNRAS.364.1105S}. In the following we will use all 15,000 available realisations of the fiducial cosmology to evaluate the quality of the various estimates.  We will take as the simulation "ground truth" the sample mean and the sample covariance based on 12,000 of these simulations while retaining 3000 simulations to test likelihoods built using the various estimators, see section \ref{sec:negLkl}.
The force mesh grid size of all the simulations is $\mathcal{N}_\mathrm{m} = 1024$.

\subsubsection{Surrogate solver}

We generate the fast surrogate samples with The COmoving Lagrangian Acceleration (COLA) method from \citet{Tassev_2013}, which allows generating approximate gravitational $N$-body outputs using a smaller number of timesteps than our simulation code.
The principle of COLA is to add residual displacements computed with a Particle-Mesh (PM) $N$-body solver to the trajectory given by analytical LPT approximations. See  \citet{Izard_2016} for comparisons of the capabilities and computational cost of COLA against $N$-body simulations in different configurations.
Like in \carpool, we used  \texttt{L-PICOLA}, a publicly available and parallel (MPI) code implementation of COLA developed by \citet{Howlett_2015}, with a force mesh grid size $N_m^{\mathrm{cola}}=512$. We computed the matter power spectra, correlation functions and PDFs using \texttt{Pylians3} \footnote{\url{https://github.com/franciscovillaescusa/Pylians3}}  and  the matter bispectra with \texttt{pySpectrum}\footnote{Available at \url{https://github.com/changhoonhahn/pySpectrum}}.

\subsection{Description of tests}\label{sec:testsDescription}
The goal of this section is to briefly explain the different tests we have implemented to assess the reliability of the CARPool covariance estimates, and to compare with the standard (bias-corrected) sample covariance matrix from simulations only.

\subsubsection{Variance reduction on matrix elements}
This test, similarly to what was assessed concerning the mean of clustering statistics in \carpool, consists in taking the empirical variance ratio, between a set of $\boldsymbol{Y}_i$ samples (vectorized outer products of centered data) and the corresponding $\boldsymbol{X}_i(\boldsymbol{\beta})$ samples, with $\boldsymbol{\beta}$ being estimated from the main set of paired simulation/surrogate samples. More precisely, in the experiments, the same set of seeds $\left\{s_i,i \in \llbracket 1,500 \rrbracket\right\}$ serves to compute both $\boldsymbol{\beta}$ and the actual covariance estimate from equation \eqref{eq:sampCARPcov}, while paired simulation/surrogate samples from seeds $\left\{s_i,i \in \llbracket 501,2000 \rrbracket\right\}$ are used to estimate $Var(\boldsymbol{Y})$ and $Var(\boldsymbol{X})$.

\subsubsection{Negative Gaussian log-likelihood on test data}\label{sec:negLkl}
Under the assumption of Gaussianity of the matter statistics, the negative log-likelihood gives a loss-function of the test data $\left\{ y_i\right\}_{test}$ from simulations. In equation \eqref{eq:mvgLkl}, the input is the covariance matrix (sample covariance or CARPool covariance) from "training" data, i.e. the seeds we reserve for covariance estimation including the "truth", while the $\boldsymbol{y}_i$ are the clustering statistics from test data simulations and $\boldsymbol{\mu}$ is replaced by $\boldsymbol{\bar{y}}$. More precisely, the sample covariance from the first 12,000 fiducial seeds of the \textit{Quijote simulations} computes the "true" covariance and gives the reference negative log-likelihood while the remaining seeds $\left\{s_i,i \in \llbracket 12001,15000 \rrbracket\right\}$ constitute the unseen test data.  

\begin{equation}\label{eq:mvgLkl}
\begin{aligned}
    -\lnb{\mathcal{L}\left( \boldsymbol{\mu}, \boldsymbol{\Sigma_{yy}}\right)}= &\frac{pN}{2}\lnb{2\pi}+\frac{N}{2}\lnb{\det\left( \boldsymbol{\Sigma_{yy}}\right)} \\
    &+ \frac{1}{2}\sum_{i=1}^{N} \left(\boldsymbol{y}_i-\boldsymbol{\mu}\right)^{\boldsymbol{T}}\boldsymbol{\Sigma_{yy}^{-1}}\left(\boldsymbol{y}_i-\boldsymbol{\mu}\right)
    \end{aligned}
\end{equation}

In section \ref{sec:results}, we show the convergence of the log-likelihood with increasing number of simulations. The loss function is computed only if the corresponding covariance estimate is positive-definite; in the numerical tests, we will see that for different statistics the CARPool covariance estimates become positive-definite for a different minimum number of samples. 

\subsubsection{Eigenspectrum}
We  compare the eigenvalues of the sample covariance from simulations with these of the CARPool covariance from simulation/surrogate pairs. It is well known that the sample covariance matrix from equation \eqref{eq:sampCov} of a vector statistics of size $p$ is at most of rank $N-1$ with $N$ samples. Moreover, as demonstrated by \citet{baiYin}, among others, the sample covariance matrix for $N\sim p$ is ill-conditioned even when full-rank: the smallest eigenvalues in particular disperse from the true values and are biased low even for unbiased estimates of the covariance matrix elements.
Therefore, computing the ratio of the ordered eigenvalues of an unbiased covariance estimate and of the "ground truth" covariance is a relevant indicator of the quality of the estimation. We stress that this test does not constitute a complete comparison of these matrices since the eigenbasis could still differ.
Still, comparing the eigenspectra of two  covariance estimates of the same random vector is common practise (e.g. \citet{2017MNRAS.466L..83J} and \citet{2008MNRAS.389..766P}) and can be considered in the context of the other, complementary tests we show.

\subsubsection{Fisher analysis and parameter uncertainties}
The covariance matrix plays a central role in parameter inference and when forecasting parameter constraints for future data sets. In this context, the relevant performance metric is not the convergence of the covariance matrix in isolation but the way parameter-driven variations in the measured quantities are constrained. This is precisely what the Fisher information matrix measures, which is why it is an essential component in parameter estimation and forecasts.

For this test we will model the likelihood of the simulated observable as a multivariate Gaussian with $\boldsymbol{y} \sim \mathcal{N} \left( \boldsymbol{\mu}(\boldsymbol{\theta}), \boldsymbol{\Sigma_{yy}}(\boldsymbol{\theta})\right)$. In this approximation\footnote{For Gaussian data whose covariance depends on the parameter values, equation \eqref{eq:fisherMVG} would include a second term. This term vanishes for the Gaussian approximation we consider here, where the statistics  are approximated as Gaussian with constant covariance around a parameter-dependent mean (see \citet{alsingwandelt2018} for a succinct explanation and \citet{carron2013} and \citet{kodwani2019} for further discussion). 
} the Fisher matrix is the symmetric matrix of size $(d,d)$ 
\begin{equation}
    \begin{aligned}
    \mathcal{F}_{ij} = \left(\frac{\partial \boldsymbol{\mu(\boldsymbol{\theta})}}{\partial \theta_i}\right)^{\boldsymbol{T}}\boldsymbol{\Sigma_{yy}^{-1}}\left(\frac{\partial \boldsymbol{\mu(\boldsymbol{\theta})}}{\partial \theta_j}\right)
    \end{aligned}\label{eq:fisherMVG}
\end{equation}

We define the vector of parameters as $\boldsymbol{\theta} = \left( \Omega_m, \Omega_b,h,n_s,\sigma_8,M_{\nu}\right)^{\boldsymbol{T}}$, and $\mathbb{E}_{\boldsymbol{\theta}}\left[\boldsymbol{y}\right] = \boldsymbol{\mu}(\boldsymbol{\theta})$ is the expectation of $\boldsymbol{y}$ for fixed parameters $\boldsymbol{\theta}$.

The Cramér-Rao inequality then gives the lower-bound of the variance of an unbiased estimator for parameter $\theta_i$, marginalized over the other parameters:
\begin{equation}\label{eq:cramerRao}
    \sigma^2_{\theta_i} \geq {\left[\mathcal{F}^{-1}\right]_{ii}}\,.
\end{equation}
The partial derivatives of the statistics are estimated numerically using finite differences from $500$  \textit{Quijote} simulations for each varying parameter exactly as in \citet{2020ApJS..250....2V} (see Table 1 of this work for the parameter values). When finite difference simulations are not already available, one can easily apply the CARPool method to the estimation of the mean of the derivatives: in \carpool, especially for the matter power spectrum and bispectrum, the precision of the CARPool mean with $5$ $N$-body simulations was comparable to that of the mean of $500$ simulations. We do not further explore this application of CARPool, since the focus of this paper is covariance estimation. 

\subsection{Results on clustering statistics covariance at $z=0.5$}\label{sec:results}

Prior to extracting clustering statistics on each snapshot, we compute the matter overdensity $\rho(r=|\boldsymbol{r}|)$ on a grid with $\boldsymbol{r}$ the comoving-coordinates in $h^{-1} {\rm Mpc}$.  The density contrast field is then $\delta(\boldsymbol{r}) \equiv \rho(\boldsymbol{r})/\bar{\rho} - 1$.
We present results at redshift $z=0.5$, which is approximately the lowest redshift that is relevant for upcoming  galaxy surveys of the large-scale structure. We found higher correlation for some statistics (power spectrum) at $z=0.0$ than $z=0.5$ and interpret that as the erasure by the non-linearities of discrepancies in the intermediate structure growth. Thus, the $z=0.5$ case may be close to the worst case and we expect CARPool to be even more efficient both for higher and for lower redshifts, either for the mean estimation like in \carpool\ or for the covariance matrix in this study. The tests described in section \ref{sec:testsDescription}, for each clustering statistics, allow examining both the covariance estimator from equation \eqref{eq:sampCARPcov} as well as its inverse, as an estimator of the precision matrix. We will assess the eigenspectrum, the negative log-likelihood loss function from equation \eqref{eq:mvCV}, and the Fisher matrix as a proxy for the adequacy of the covariance matrix estimate for deriving parameter constraints. We also show the element-by-element variance reduction on the new estimate (performance of CARPool) and plot the covariance and precision matrices to provide a visual cue of the reduction in noise with respect to the sample covariance. As discussed in section \ref{sec:appcov}, the precision estimate $\boldsymbol{\Sigma_{yy}^{-1}}$ will not include the Hartlap factor, whether for the sample covariance or the CARPool covariance.

\subsubsection{Matter Power Spectrum}\label{sec:pk}
The density contrast $\delta(\boldsymbol{x})$ is computed on a square grid of size $N_\mathrm{grid} = 1024$ for each snapshot, then, in 3D Fourier space, the average of $|\delta\left(k\right)|^2, k \in \left[ k-\Delta k, k+\Delta k\right]$ gives the power spectrum $P(k)$ for wave vector modulus $k$. The \textit{Quijote} power spectra range from $k_\mathrm{min}=$ \num{8.900e-3} $h {\rm Mpc^{-1}}$ to $k_\mathrm{max}=5.569$ $h {\rm Mpc^{-1}}$.
The following analysis is restricted between $k_\mathrm{min}=$ \num{8.900e-3} $h {\rm Mpc^{-1}}$ and $k_\mathrm{max}\approx 1.0$ $h {\rm Mpc^{-1}}$, which results in $p=79$ linearly spaced bins. Therefore, we have $P=3160$ unique covariance matrix elements to estimate.

\begin{figure*}
    \includegraphics[width=\textwidth]{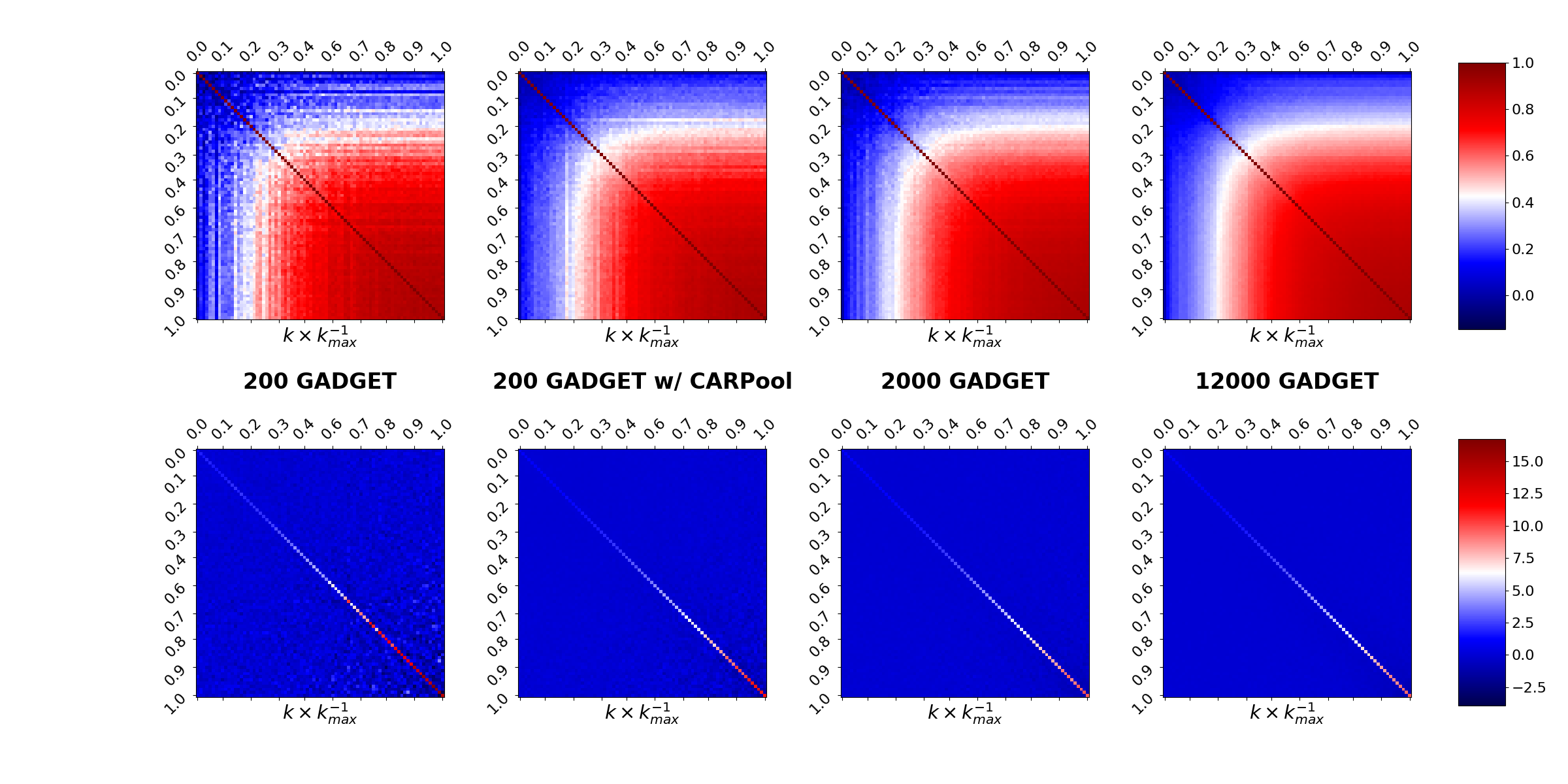}
    \caption{We display multiple power spectrum covariance estimates (top) and their inverse (bottom) in order to illustrate the improvement on the standard sample covariance matrix. For reference, the power spectrum is shown on the same axes in Figure \ref{fig:empVarCovPk}. Covariance matrices are shown as their correlation counterpart $\boldsymbol{D}
    ^{-1}\boldsymbol{\widehat{\Sigma}}\boldsymbol{D}
    ^{-1}$ with the diagonal $\boldsymbol{D} = \sqrt{\mathrm{diag}\left( {\boldsymbol{\widehat{\Sigma}}}\right)}$. The precision matrices at the bottom are the inverse of the corresponding correlation matrix at the top.}
    \label{fig:visualPk}
\end{figure*}

\begin{figure}
    \includegraphics[width=\columnwidth]{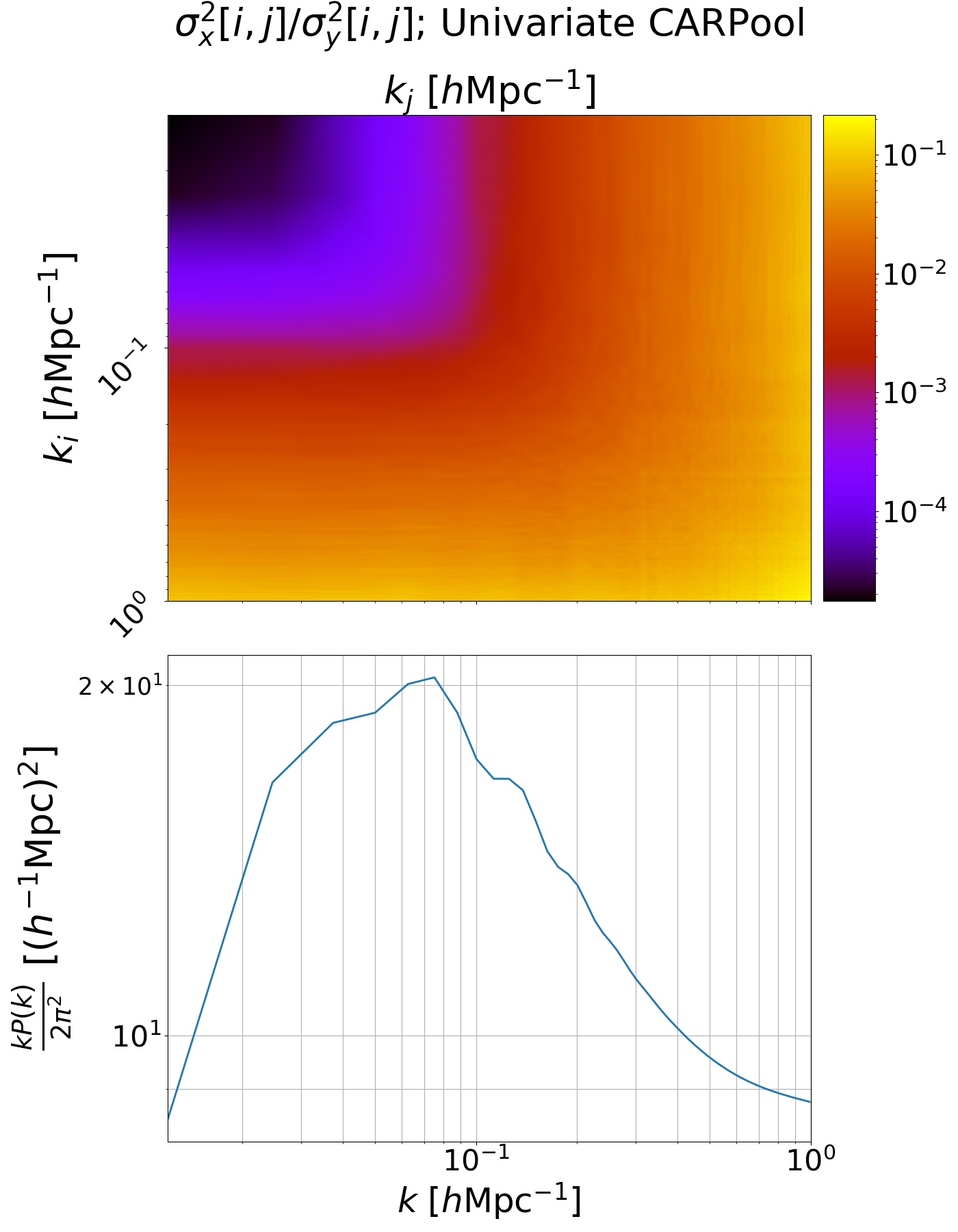}
    \caption{Top: We show the variance reduction with CARPool of the estimated power spectrum second-order moments up to $k_\mathrm{max} \approx 1.0~h {\rm Mpc^{-1}}$. The control matrix that generates 3,200 $\boldsymbol{X}_n$ CARPool samples is estimated with $200$ paired realisations.
    The variance of the covariance elements of $\boldsymbol{\Sigma_{yy}}$ are estimated using 3,200 available power spectra from the \textit{Quijote} simulations.
    Bottom: For reference, reduced power spectrum from the mean of the \textit{Quijote} simulations at $z=0.5$.}
    \label{fig:empVarCovPk}
\end{figure}

\begin{figure}
    \includegraphics[width=\columnwidth]{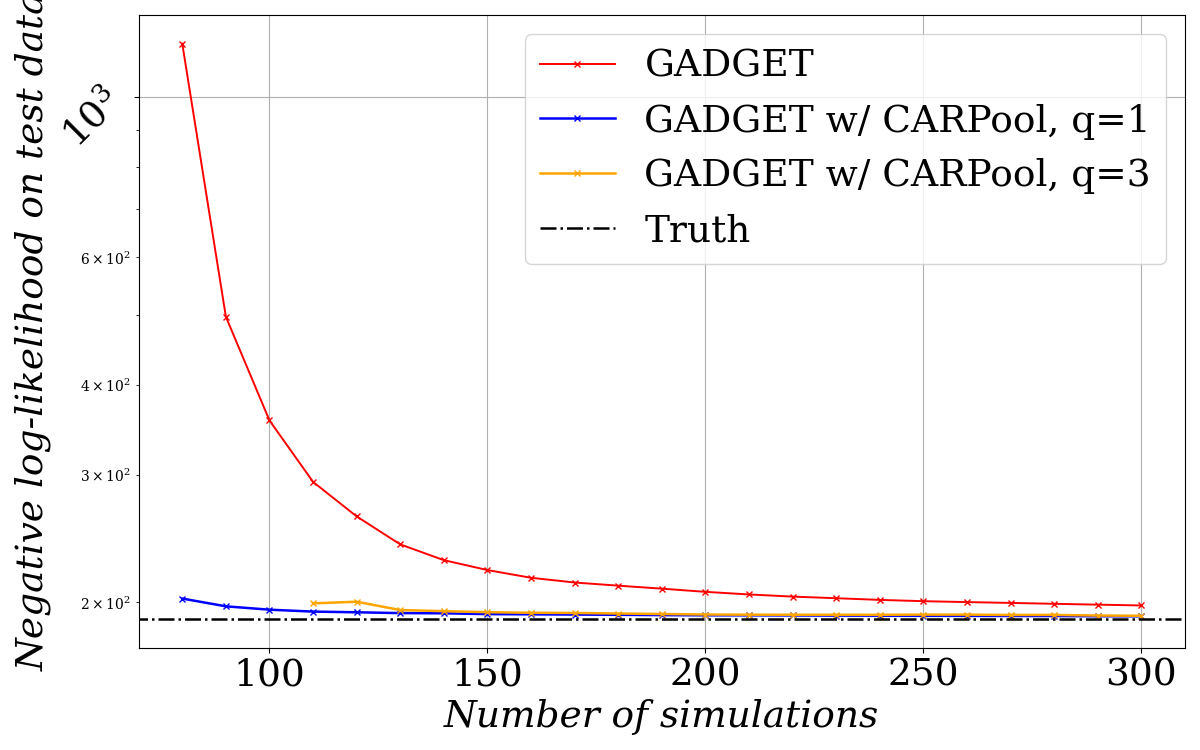}
    \caption{Negative log-likelihood on test data -- acting as a cost function -- evaluated for an increasing number of available $N$-body simulations used to estimate the covariance matrix of the matter power spectrum. We observe that the CARPool estimates converge much faster towards the true value of the cost function.}
    \label{fig:neglogPk}
\end{figure}

\begin{figure}
    \includegraphics[width=\columnwidth]{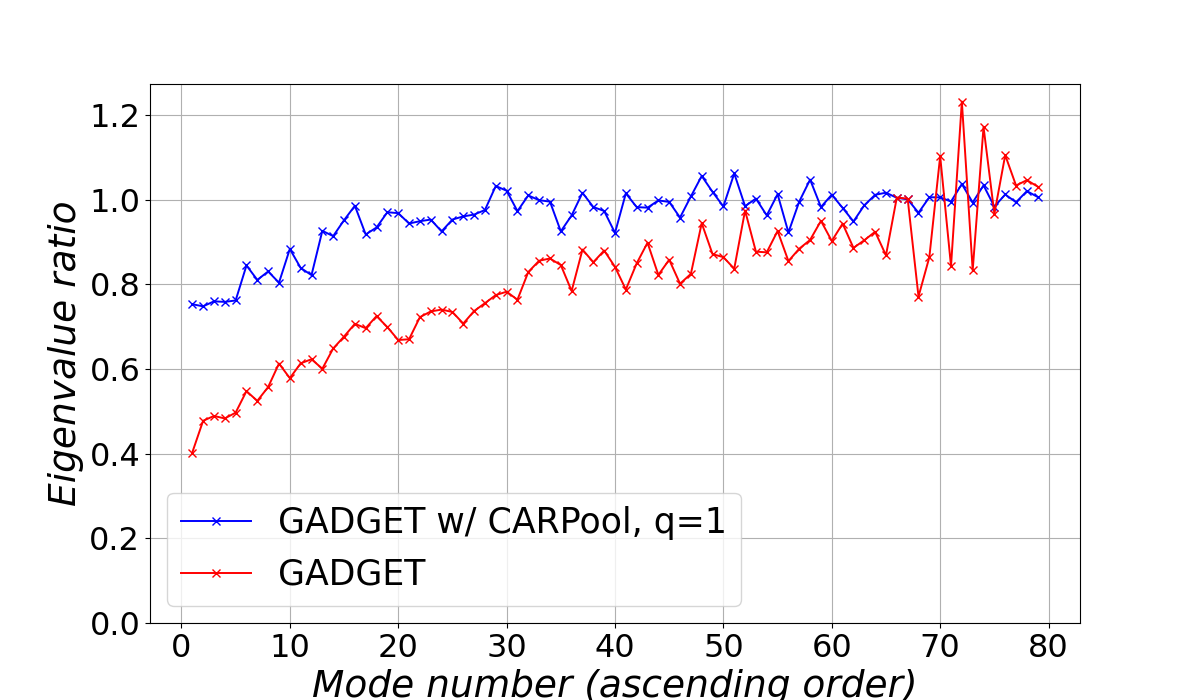}
    \caption{We show the improvement on the conditioning of the covariance estimate with CARPool by showing the ratio of the eigenvalues in ascending order between the estimated covariance using $200$ simulations and the "true" covariance matrix estimated with $12000$ simulations: $\lambda_i^{test}/\lambda_i^{true}$ for each index $i$. A constant line at 1 would indicate identical eigenspectra but would not imply that the eigenbases are the same, as discussed in section \ref{sec:testsDescription}}.
    \label{fig:eigenPk}
\end{figure}

Figure \ref{fig:visualPk} shows the CARPool estimate of the covariance and the precision matrix using $200$ simulations. For comparison, we show the sample covariance estimates for $200$ and $2000$ simulations as well as the "ground truth" covariance measured from 12,000 \texttt{GADGET}  simulations. The empirical variance reduction on each estimated covariance matrix element appears in Figure \ref{fig:empVarCovPk}: as expected, the variance reduction on the $\boldsymbol{X}(\boldsymbol{\beta})$ samples at large scales is much higher ($\sim 10^4$) than for variances and cross-covariances at small scales ($\sim 10$-fold reduction). In Figure \ref{fig:neglogPk}, we see that the $q=1$ case is positive definite from $N=80$ simulations onward while with $q>1$ (we stopped at $q=3$ for the power spectrum) attains positive definiteness at$\sim 120$ simulations and offers no improvement on the loss function on test data, at least for a small to moderate number of simulations and for our choice of neighbourhood induced by the vectorization of the covariance matrix, \textit{c.f} Figure \ref{fig:vectorization}. The log-likelihood of the CARPool covariance   converges much faster to the log-likelihood of the "ground truth" covariance than that of the sample covariance based on the same number of simulations.  Additionally, Figure \ref{fig:eigenPk} demonstrates that for the same number of simulations ($N=200$ in this plot), the eigenvalue ratio greatly favors the CARPool estimate: small eigenvalues, which are the last to converge when using the sample covariance, are lifted up and the largest modes are more stable. 

\begin{figure*}
    \includegraphics[width=\textwidth]{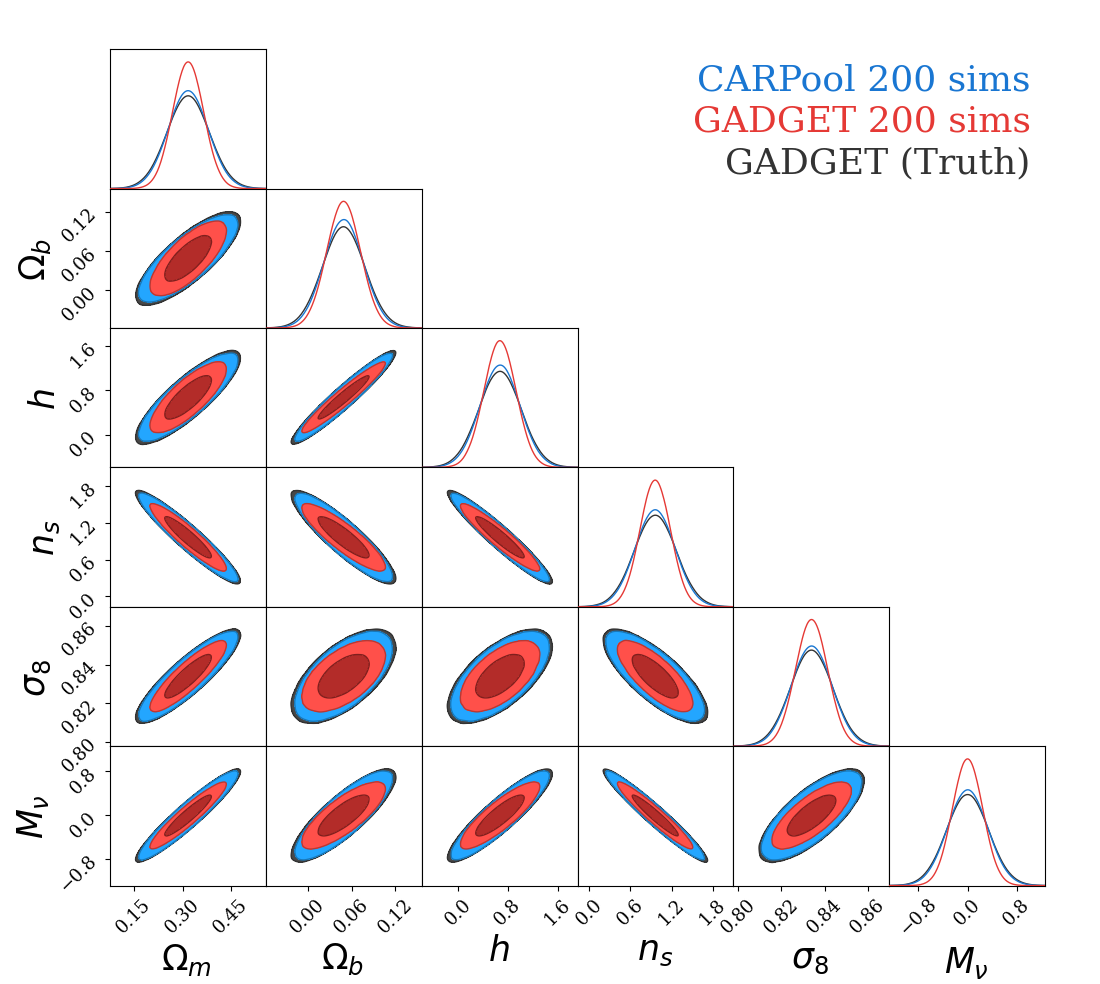}
    \caption{Confidence contours of the cosmological parameters computed using the Fisher matrix based on the estimated matter power spectrum covariance matrix.  The "truth" designates the confidence region using the sample covariance matrix of 12,000 $N$-body simulations, and the mean parameters are known from the $\Lambda$CDM models used in the simulations. We thus demonstrate the better conditioning of the CARPool covariance estimate with respect to the sample covariance for the same number of simulations.}
    \label{fig:fisherPk}
\end{figure*}

How does this improvement in the covariance matrix translate to the Fisher matrix?  We can see in Figure \ref{fig:fisherPk} that, with respect to the Fisher matrix computed using the "ground truth" covariance the sample covariance of size $(79,79)$ using $200$ simulations  leads to a significant underestimate and, in some cases, rotation of the confidence regions of the parameters. The CARPool covariance using the same number of simulations (plus the paired and additional surrogate samples) gives a considerably more accurate Fisher matrix. 

\subsubsection{Matter Bispectrum}
In this subsection we turn to estimating the covariance of the matter bispectrum. Like in \carpool, we will consider two separate subsets of the matter bispectrum: the matter bispectrum monopole $B(k_1,k_2,k_3)$ of squeezed isosceles triangles on the one hand, and the reduced bispectrum monopole $Q(k_1,k_2,k_3)$ for equilateral configurations on the other hand. We will apply the same tests we used for the  power spectrum covariance matrix, looking at the variance reduction,  the negative log-likelihood, the eigenspectrum, and the Fisher matrix computed from the estimated bispectrum covariance matrix.

\paragraph{Squeezed isosceles triangles}
We build the first group of samples by grouping triangle configurations for which $k_1=k_2$ and by ordering the bispectrum monopoles in ascending order of the $k_3/k_1$ ratio. We keep squeezed triangles: $\left(k_3/k_1\right)_\mathrm{max} = 0.20$ ($p=98$ and $P=3851$). Since $q=3$ gives a slight improvement over $q=1$ the figures will show the results for $q=3$.

\begin{figure*}
    \includegraphics[width=\textwidth]{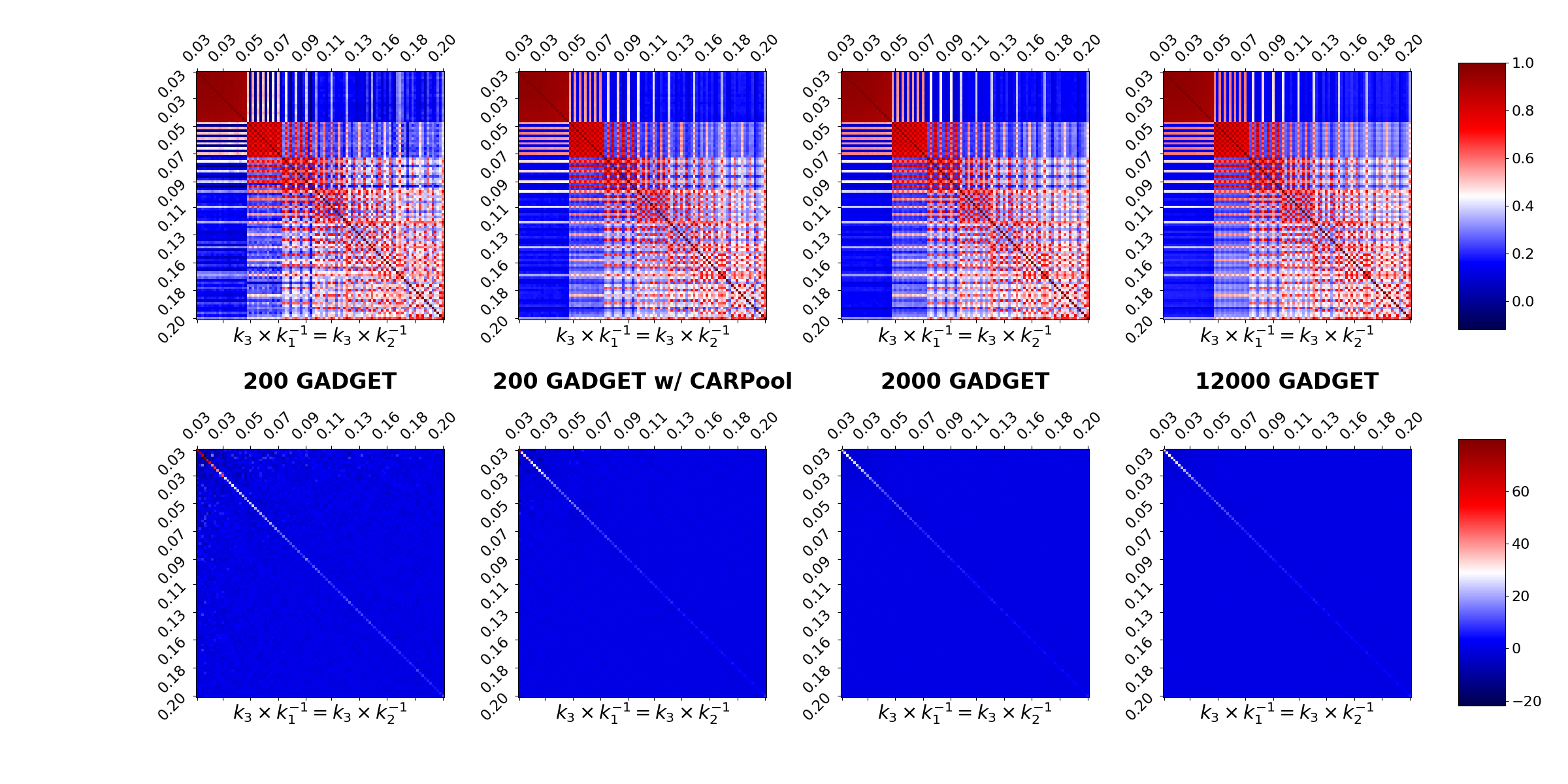}
    \caption{We plot different matter bispectrum covariance estimates (top) and their inverse (bottom), for squeezed isosceles triangles, similarly to Figure \ref{fig:visualPk}.}
    \label{fig:visualBkSq}
\end{figure*}

\begin{figure}
    \includegraphics[width=0.49\textwidth]{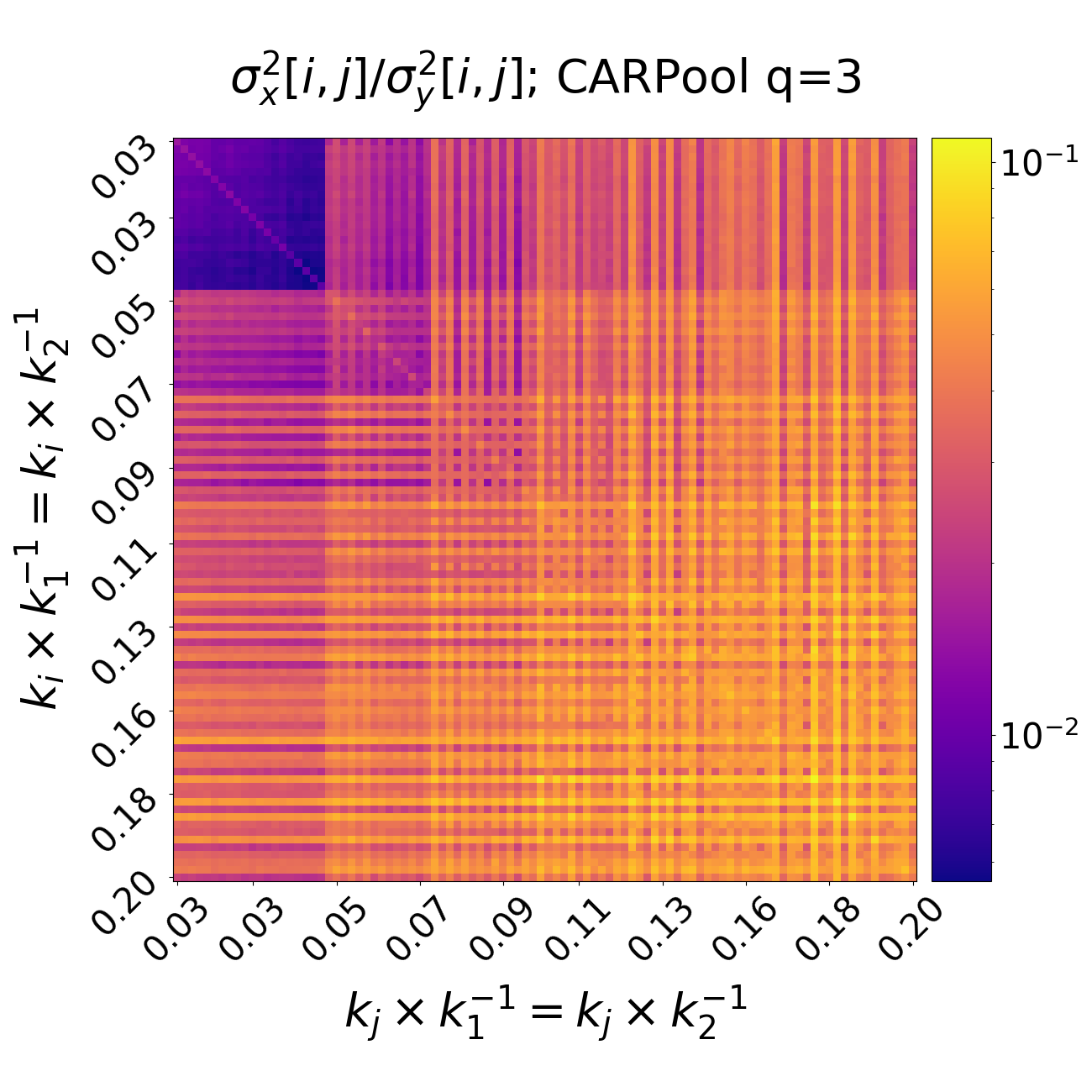}
    \caption{We demonstrate there is significant variance reduction of the estimated matter bispectrum covariance matrix elements, for the set of squeezed isosceles triangles up to $\left(k_3/k_1\right)_{max}=0.2$. The control matrix that generates 1,800 new $\boldsymbol{X}_n$ CARPool samples, to compute the variance of each vector element, is estimated with $200$ paired realisations.
    To estimate the variance of each element of $\boldsymbol{\Sigma_{yy}}$, we use 1,800 samples from \textit{Quijote} simulations.}\label{fig:empVarCovBkStd}
\end{figure}

Figure \ref{fig:visualBkSq} compares the CARPool covariance estimate with the sample covariance  estimator applied 200, and 2000 \texttt{GADGET} simulations with the "ground truth" computed from 12000 simulations. The differences are visually most apparent in the precision matrix, where the CARPool estimate from $200$ simulations looks visually similar to the standard estimate from $200$ simulations. Figure \ref{fig:empVarCovBkStd} gives a quantitative view of the CARPool variance reduction of the covariance matrix elements. As for the power spectrum covariance, CARPool also improves the eigenvalues of the covariance matrix (Figure \ref{fig:eigenBkSq}).

\begin{figure}
    \includegraphics[width=0.49\textwidth]{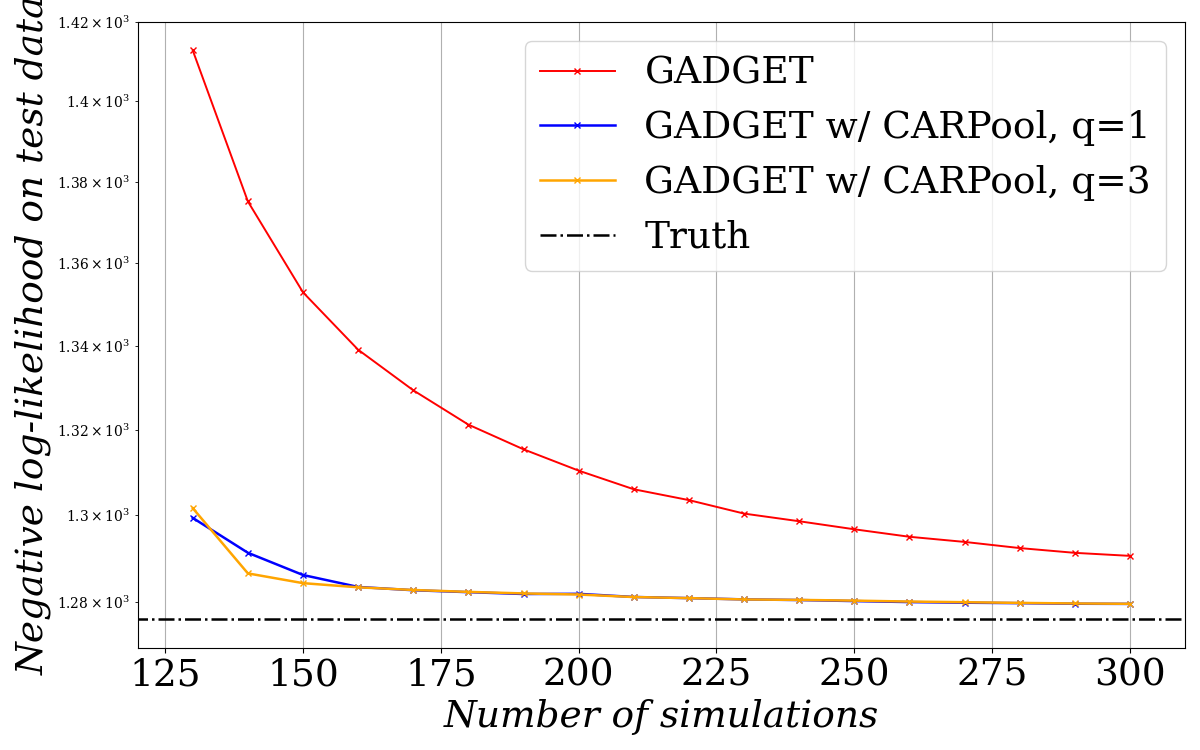}
    \caption{Negative log-likelihood on test data evaluated for an increasing number of simulations used to compute a covariance estimate, like in Figure \ref{fig:neglogPk}}\label{fig:negLHBkSq}
\end{figure}

\begin{figure}
    \includegraphics[width=\columnwidth]{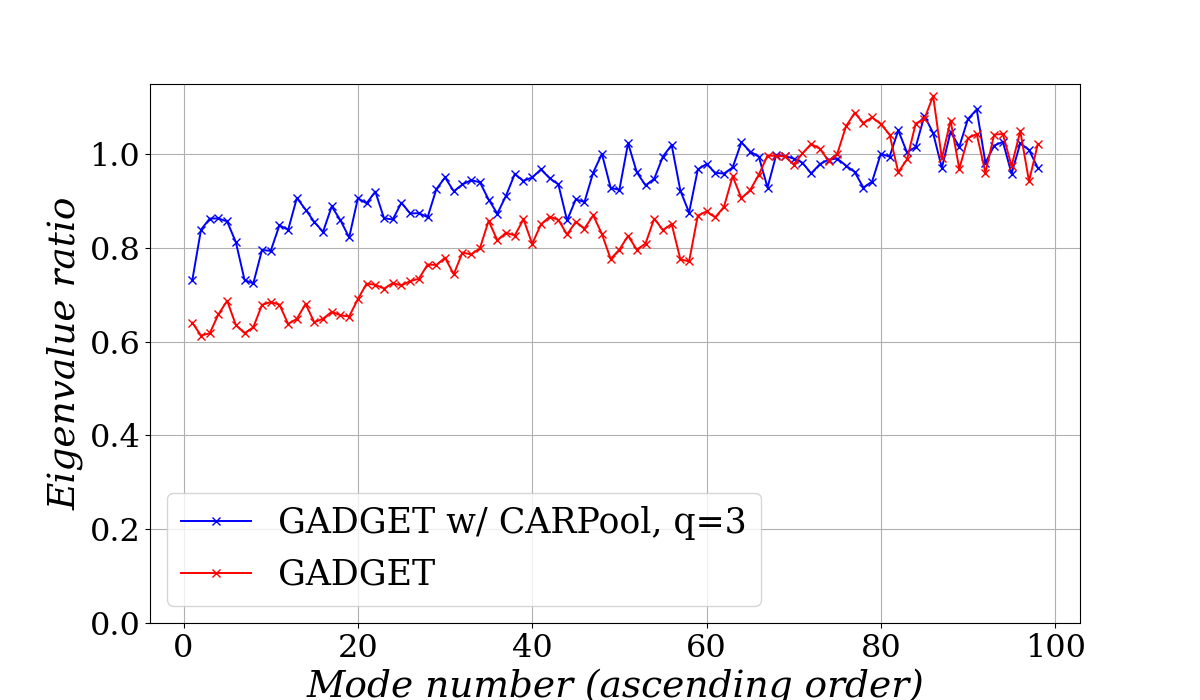}
    \caption{The computation method for the eigenvalue ratio of the matter bispectrum covariance is identical to Figure \ref{fig:eigenPk}}
    \label{fig:eigenBkSq}
\end{figure}

For this case, the convergence of $q=1$ and $q=3$ CARPool estimates  in term of negative log-likelihood of test data appears in Figure \ref{fig:negLHBkSq}.  We also tested $q=5$, but in that case the added noise in the estimate of $\boldsymbol{\beta}$, now a $(5,5)$ matrix, for each empirical counterpart of equation \eqref{eq:hybridBeta} worsens the performance if we limit ourselves to a moderate number of simulations.  

\begin{figure*}
    \includegraphics[width=\textwidth]{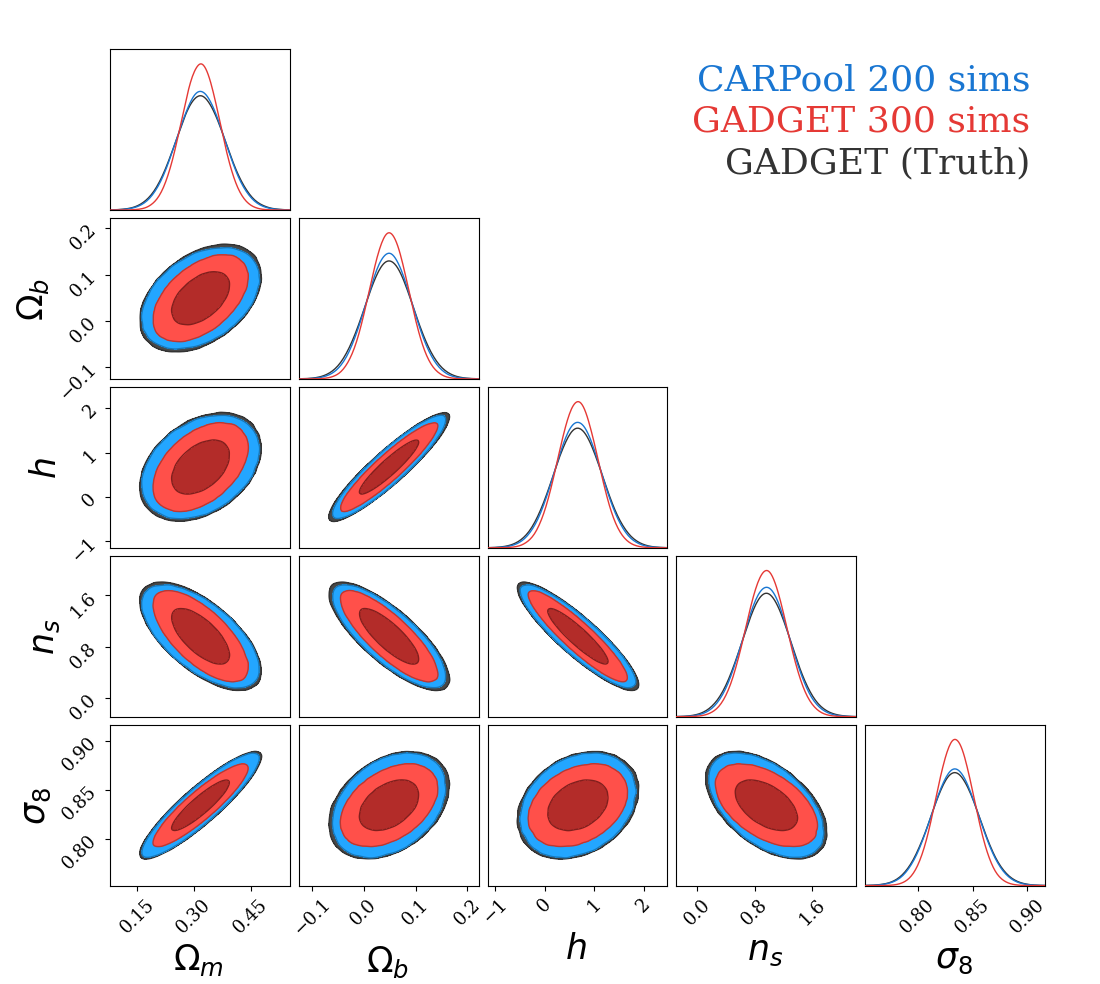}
    \caption{Confidence contours of the cosmological parameter computed using the Fisher matrix, based on the estimated squeezed matter bispectrum covariance matrix. Again, the CARPool covariance gives more realistic confidence regions than the sample covariance for an equivalent number of simulations. The "truth" designates the confidence region using the sample covariance matrix of 12,000 $N$-body simulations, and the mean parameters are known from the $\Lambda$CDM models used in the simulations.}
    \label{fig:fisherBkSq}
\end{figure*}

Figure  \ref{fig:fisherBkSq} shows that the Fisher matrix computed using the CARPool covariance based on $200$ simulations is much more accurate even than the sample covariance using  $300$ simulations, where confidence contours are underestimated compared to the "ground truth" reference computed based on 12,000 simulations\footnote{We did not include $M_{\nu}$ here; the fiducial realisations of the bispectrum initialized with the Zel'dovich approximation that are needed to compute the neutrino mass derivative as in \citet{2020ApJS..250....2V} were not available.}.

\begin{figure*}
    \includegraphics[width=\textwidth]{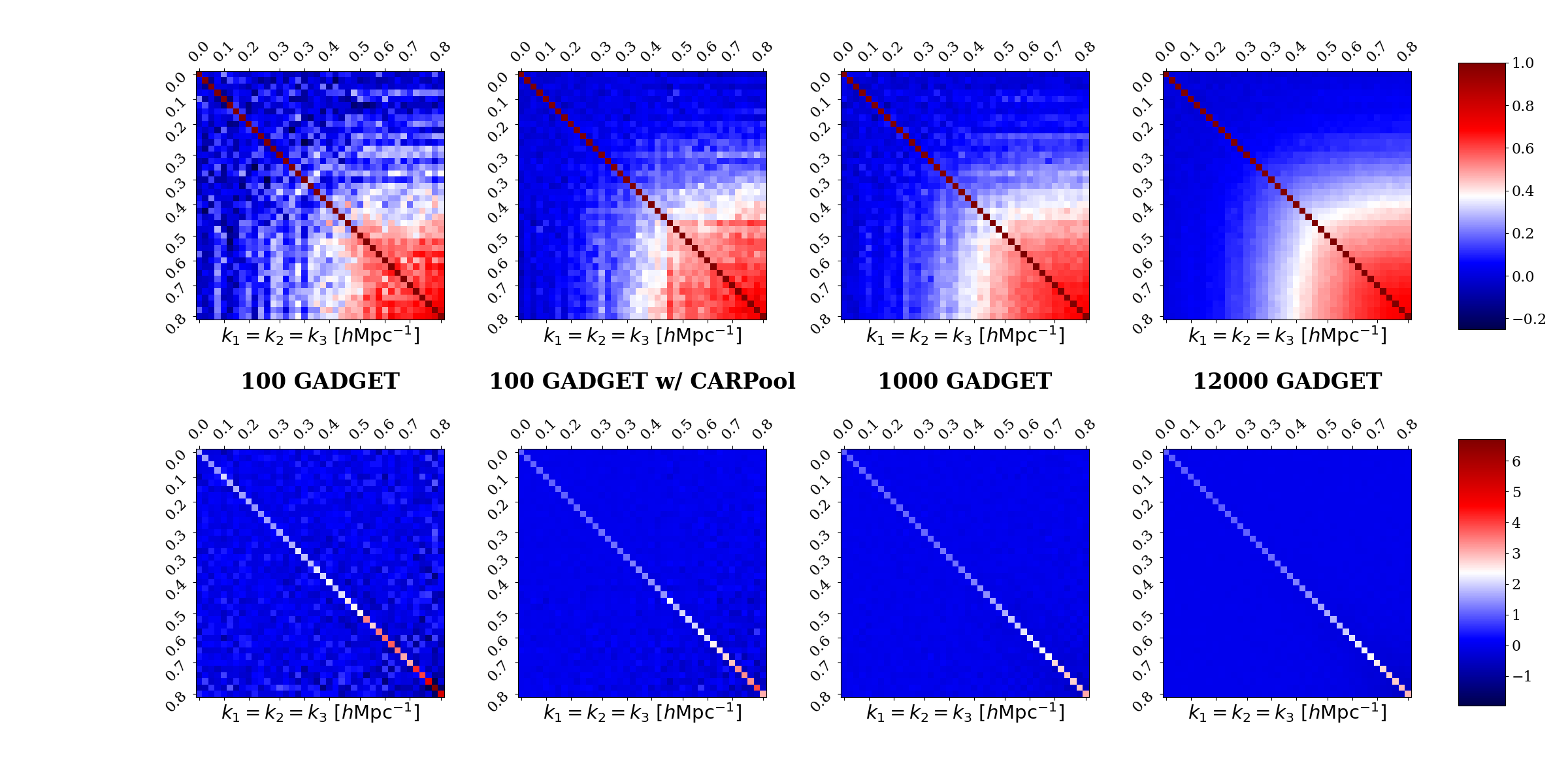}
    \caption{We plot covariance and precision matrices estimates for the reduced bispectrum of equilateral triangles, similarly to Figure \ref{fig:visualPk}}
    \label{fig:visualBkEq}
\end{figure*}

\begin{figure}
    \includegraphics[width=\columnwidth]{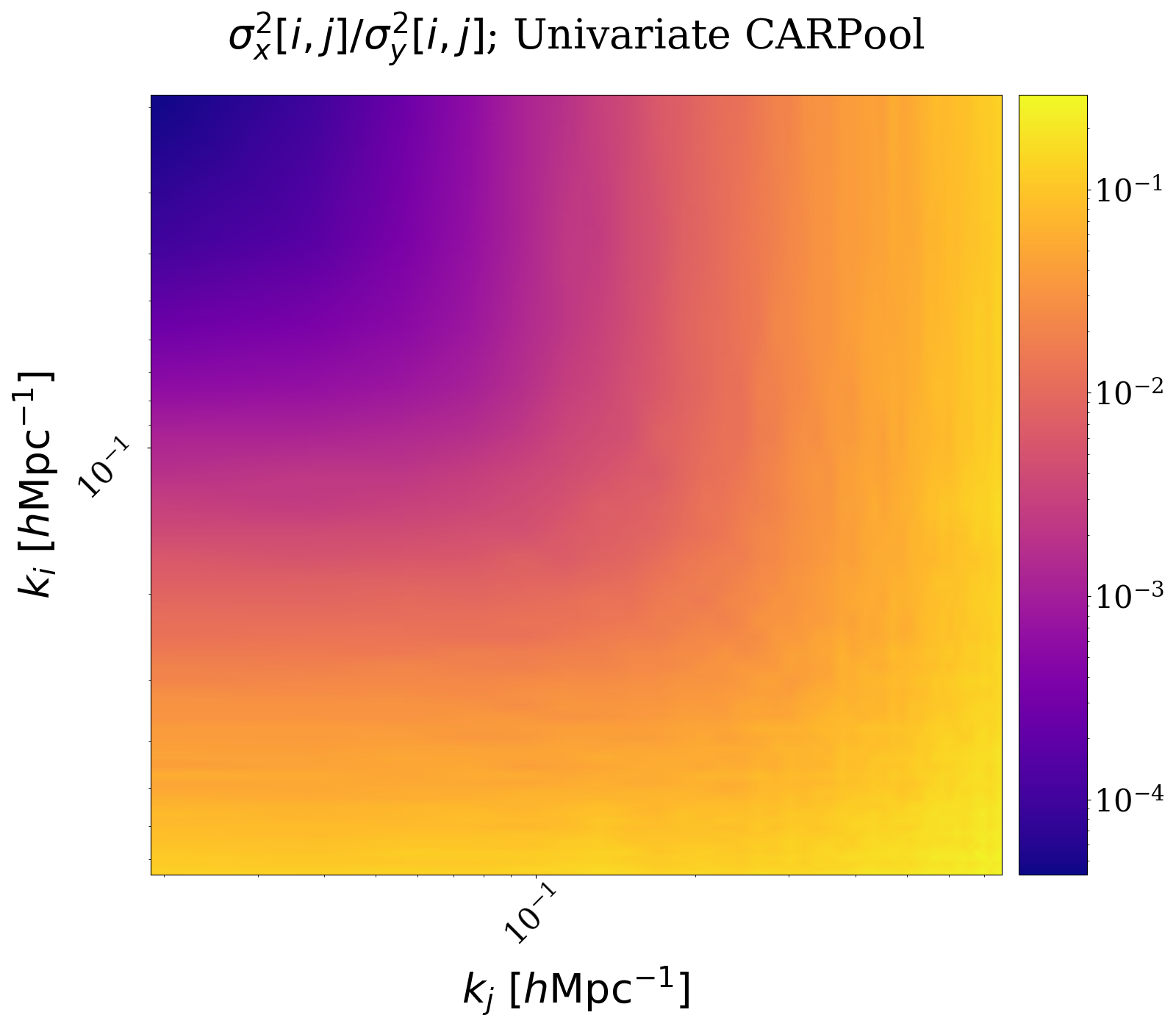}
    \caption{We exhibit the significant variance reduction for the estimated matter reduced bispectrum covariance matrix elements, for the set of equilateral triangles up to $k_1=k_2=k_3 \approx 0.75$. The computation of the metric is identical to Figure \ref{fig:empVarCovBkStd}}.
    \label{fig:empVarCovBkEq}
\end{figure}

\begin{figure}
    \includegraphics[width=0.49\textwidth]{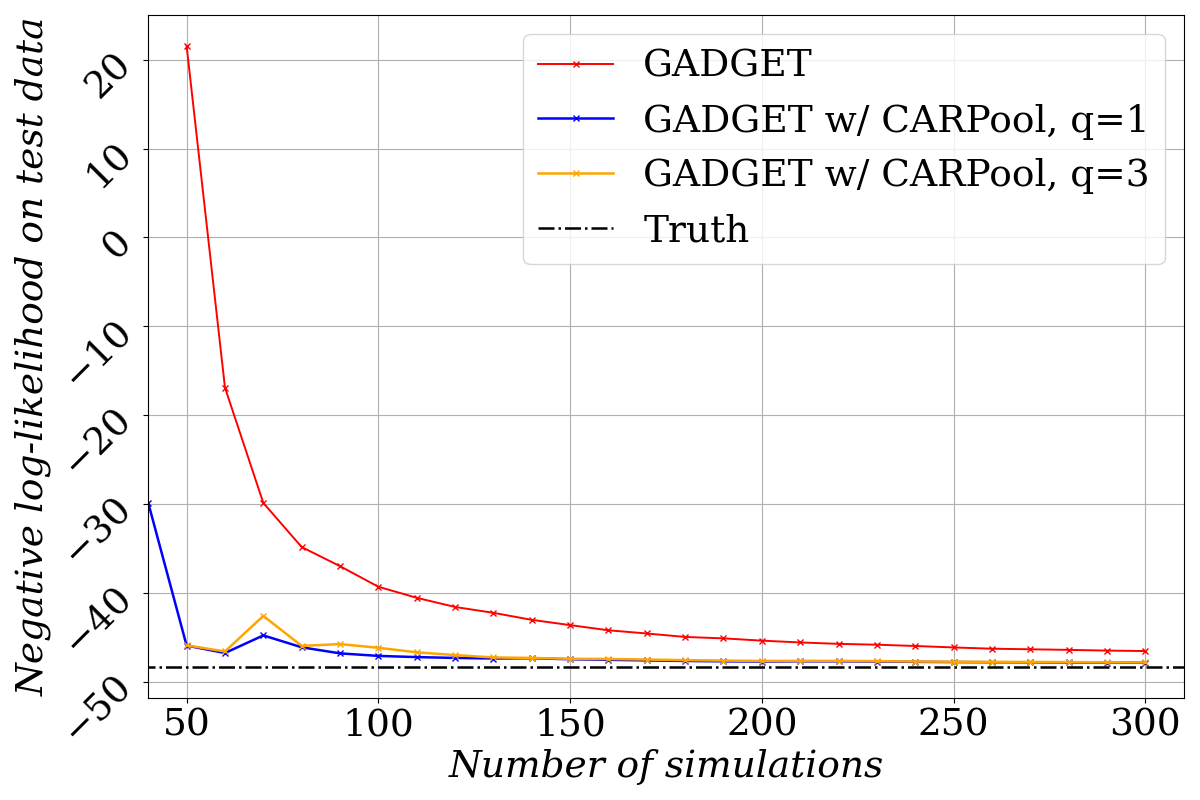}
    \caption{Negative log-likelihood on test data, evaluated for an increasing number of available $N$-body simulations that intervene in the estimation of the covariance matrix of the reduced matter bispectrum.}\label{fig:negLHBkEq}
\end{figure}

\begin{figure}
    \includegraphics[width=\columnwidth]{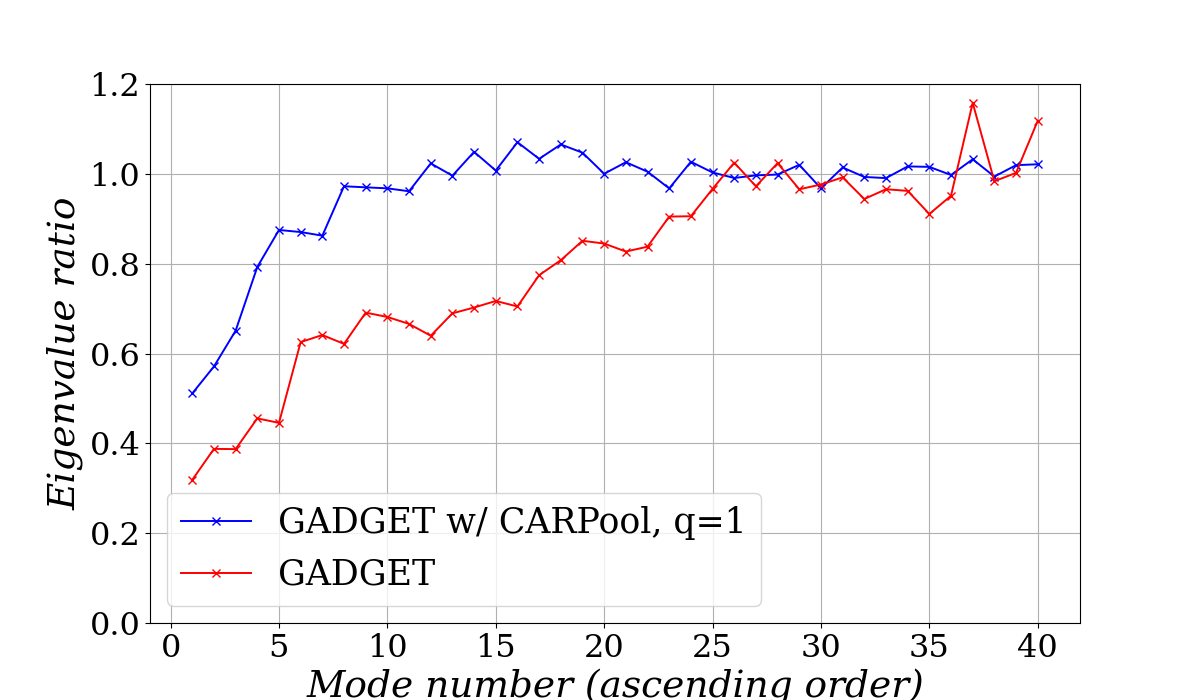}
    \caption{Same as Figure \ref{fig:eigenPk} but for the reduced matter bispectrum of equilateral triangles covariance; note that in this case only $100$ simulations were used in both the standard and the CARPool estimates.}
    \label{fig:eigenBkEq}
\end{figure}

\begin{figure*}
    \includegraphics[width=\textwidth]{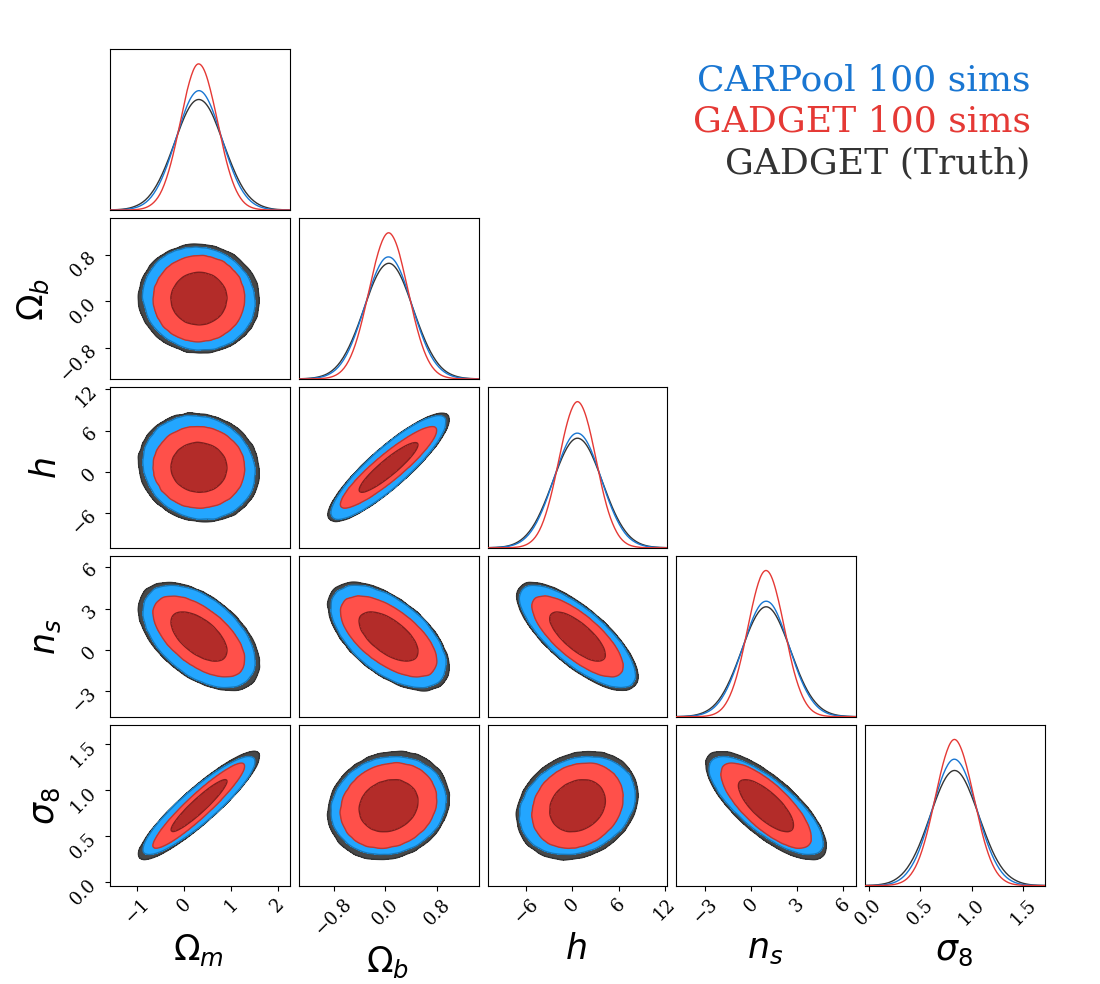}
    \caption{Same as Figure \ref{fig:fisherBkSq}; but based on the estimated equilateral matter bispectrum covariance matrix.}
    \label{fig:fisherBkEq}
\end{figure*}

\paragraph{Equilateral triangles}
The second set of bispectrum statistics is comprised of equilateral triangles with  $k_1=k_2=k_3$ varying up to $k_\mathrm{max} = 0.75$ $h{\rm Mpc^{-1}}$ ($p=40$ and $P=820$). CARPool gives a particularly strong variance reduction for this case with a smaller $p$ than before, so we focus on a case with only $100$ simulations. Figure \ref{fig:visualBkEq} visually compares the covariance matrix estimators. Figure \ref{fig:empVarCovBkEq} shows the strong variance reduction on the covariance matrix elements from $\sim \mathcal{O}(10^4)$ at large scales down to $\sim \mathcal{O}(10)$ at small scales. Figure \ref{fig:negLHBkEq} emphasizes the improvements of the log-likelihood test, with the simplest, diagonal ($q=1$) CARPool estimator being favoured  ($10$ more simulations than the sample covariance are required for that in the $q=3$ case). The eigenvalue ratios for the CARPool estimate, in Figure \ref{fig:eigenBkEq}, approach the "ground truth" even with only $100$ $N$-body simulations, except for 7 smallest eigenvalues. The Fisher analysis presented in Figure \ref{fig:fisherBkEq} exhibits the same behavior as for the previous statistics: the sample covariance with few simulations underestimates parameter confidence intervals, while the CARPool covariance with few simulation is much more representative of the knowledge about parameters given by the clustering statistic of interest. We note however, that the set of $40$ equilateral triangles we treated is much less informative about the parameters than the other statistics we consider in this paper.

\subsubsection{Matter correlation function}

\begin{figure}
    \includegraphics[width=\columnwidth]{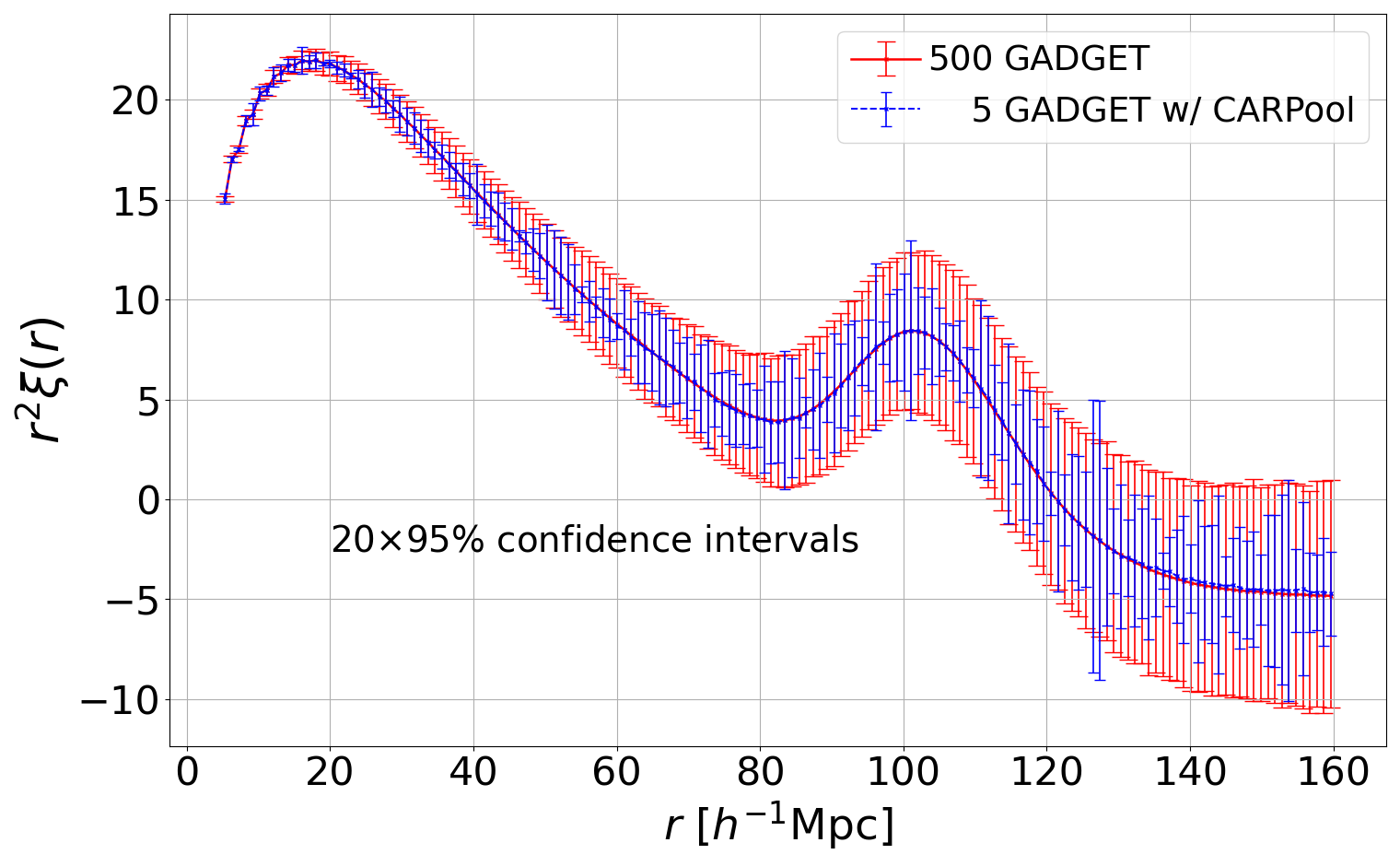}
    \caption{Estimated mean of the matter correlation function with $500$ $N$-body simulations versus $5$ pairs of ``$N$-body + cheap'' simulations. The estimated $95\%$ confidence intervals are computed with the Student $t$-score for CARPool bias-corrected and accelerated (BCa) bootstrap for the sample mean of simulations only.} \label{fig:meanCorr}
\end{figure}

We also tested real-space clustering statistics, the first of which being the two-point matter correlation function $\xi(\boldsymbol{r})$ for $\boldsymbol{r} \in \left[5.0, 160.0\right] h^{-1} {\rm Mpc}$ ($p=159$)
As we did not experiment with the correlation function in \carpool, we show the reduction of variance for the estimation of the mean in Figure \ref{fig:meanCorr}. With $5$ $N$-body simulations and CARPool, we get an unbiased estimate of the mean correlation function with an equivalent precision -- in terms of $95\%$ confidence intervals -- as with the sample mean of $500$ simulations.

\begin{figure}
    \includegraphics[width=\columnwidth]{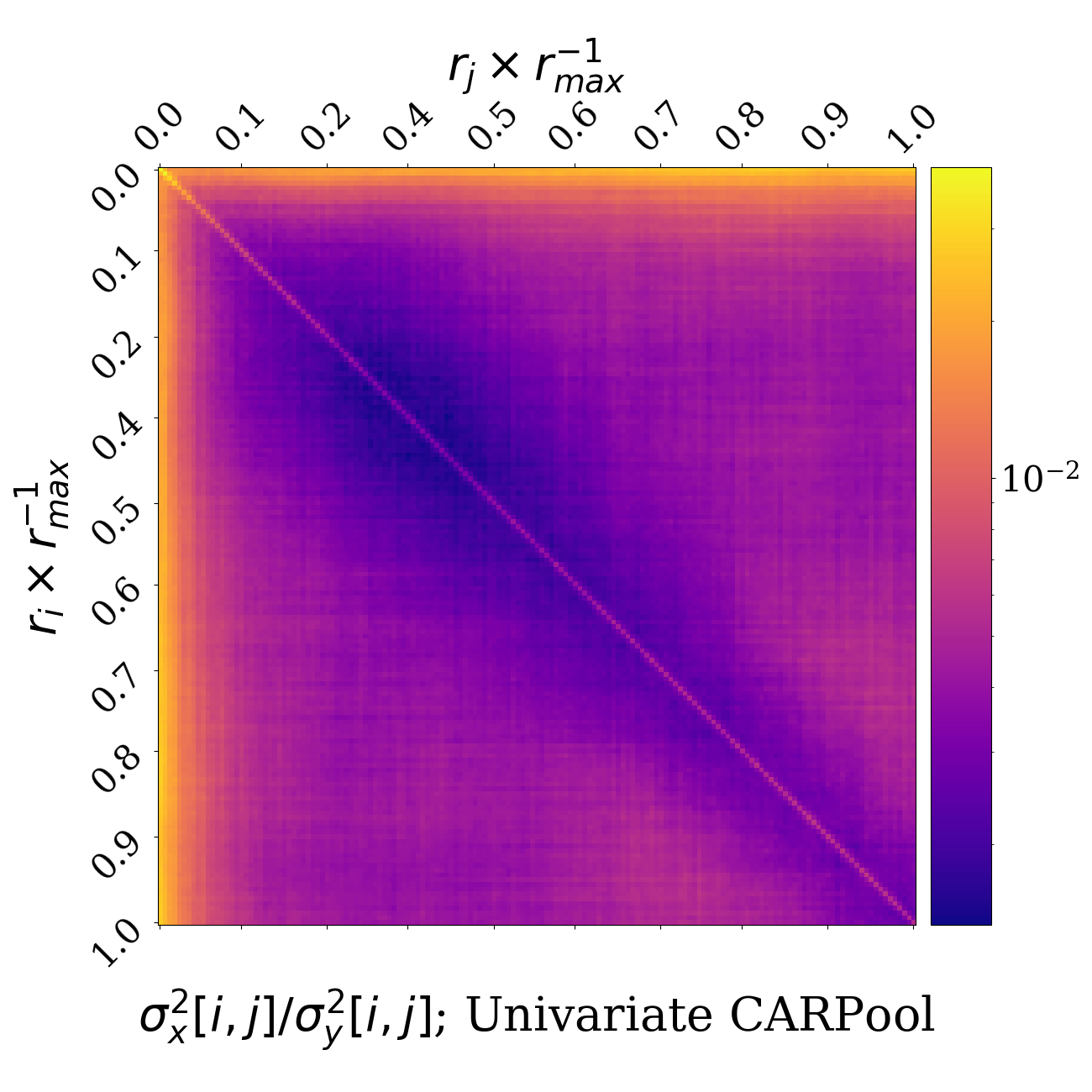}
    \caption{Variance reduction for the matter correlation function covariance matrix elements, for $r \in \left[ 5.0, 160.0\right]$ $h^{-1} {\rm Mpc}$. The computation is similar to that of Figure \ref{fig:empVarCovPk}
     } \label{fig:empVarCorr}
\end{figure}

Then, regarding the covariance matrix, we show in in Figure \ref{fig:empVarCorr} that we get consequential variance reduction on all the estimated second-order moments of the correlation function vector. We note than the reduction is "homogeneous" in the matrix, since the highly-correlated large-scale modes in Fourier space intervene at all scales in real-space by summation.

We did not however get significant improvement on the conditioning of the covariance matrix with CARPool comparatively to the sample covariance. In other words, CARPool does its job for the matter correlation function -- that is to say reducing variance on covariance elements -- but this improvement did not translate into a better eigenspectrum or strong improvements of the Fisher matrix contours.

\subsubsection{Matter PDF}
\label{sec:matterPDF}
As in \carpool, the matter PDF is computed on a grid with $N_\mathrm{grid}=512$ and smoothed by a top-hat filter of radius $R=5~h^{-1} {\rm Mpc}$. We have the raw $100$ histogram bins in the range $\rho/\bar{\rho} \in \left[ 10^{-2}, 10^{2}\right]$. Given that the covariance for this case is formally degenerate since the histogram bins are linearly dependent---each bin can be written as 1 minus the sum of the others---we have taken all the bins that are non-zero across all samples up until the tails and down-sampled the PDF by a factor $2$, which gives $p=33$ bins. Even after this modification the covariance is still nearly degenerate. There are also strong bin-to-bin correlations that suggest going to even coarser binning would improve the condition of the matrix; we proceeded without processing to test the CARPool covariance estimate in this regime.

\begin{figure}
    \includegraphics[width=\columnwidth]{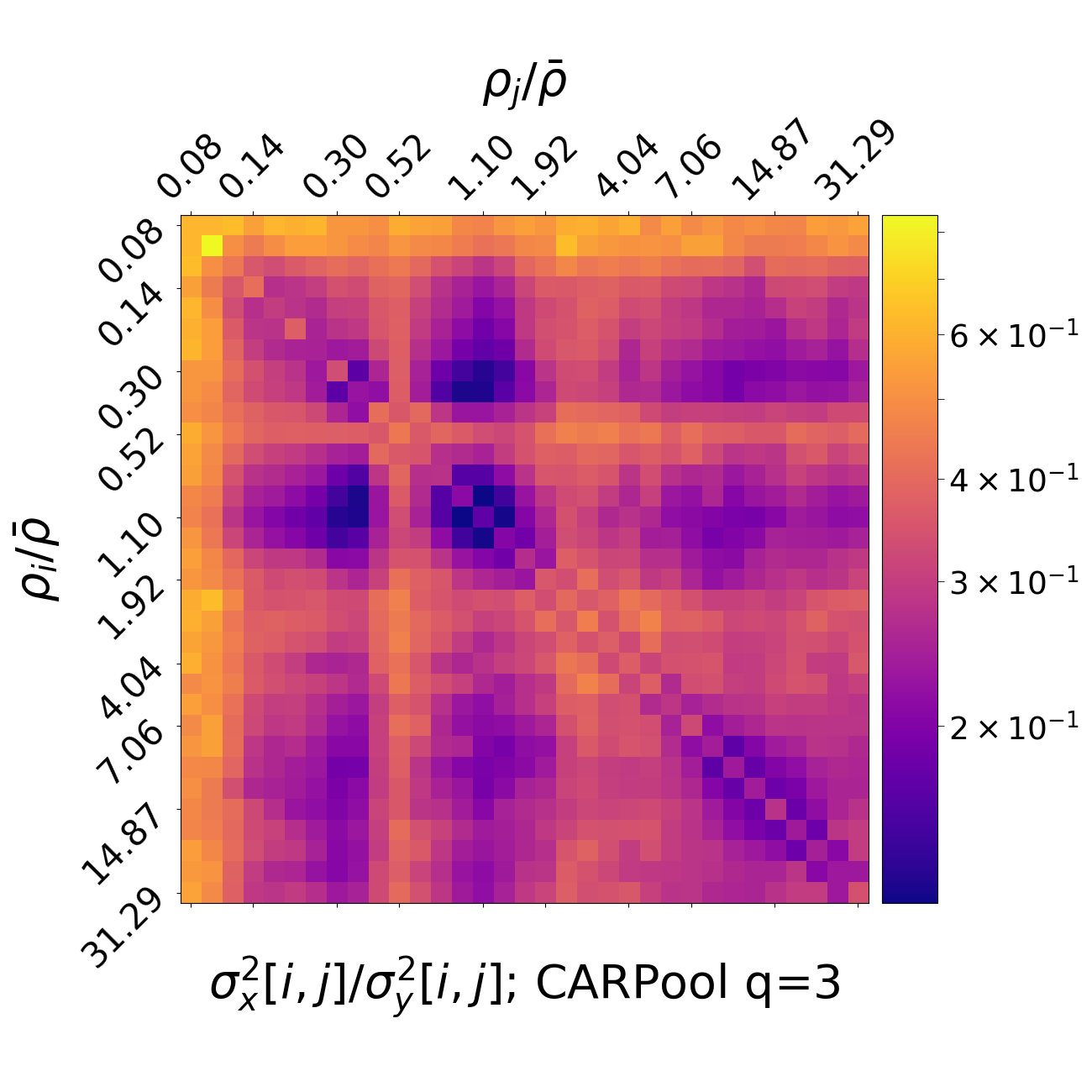}
    \caption{Variance reduction for the one dimensional matter probability density function, down-sampled by a factor $2$. The computation is similar to that of Figure \ref{fig:empVarCovPk}
     }
    \label{fig:empVarCovPDF}
\end{figure}

Figure \ref{fig:empVarCovPDF} shows that the variance of the CARPool covariance estimate is only mildly reduced. A similar effect was seen in the CARPool estimate of the mean PDF. Since the densities of structures in the COLA surrogates do not match the densities of the corresponding structures in the simulations, underestimating the density in halos and overestimating underdensities in voids, fluctuations in density bins of the surrogate are correlated to fluctuations in other density bins of the simulation. A larger variance reduction would be obtained with a brute-force dense control matrix $\boldsymbol{\beta}$, though this would require a large number of simulations to estimate the control matrix and thus defeat the point of the approach. An alternative would be to define a pre-processing function to map the average density PDFs of the surrogates to approximately match the average PDF of the simulation, as in \cite{2013JCAP...11..048L}. This would likely increase the bin-to-bin correlations for diagonal control matrix  and therefore improve the CARPool estimates. We did not pursue these ideas further in order to test the CARPool approach without designer pre-processing.

We found that as a consequence of the small variance reduction and the near-degeneracy of the covariance, the eigenspectrum of the resulting covariance estimates is not improved and in fact not positive-definite for all test realisations. We will discuss possible remedies to this issue below.

\section{Discussion and Conclusion}
\label{sec:conclusions}
We explore the problem of estimating covariance matrices as a new application of the CARPool principle, \textit{i.e.}, of combining simulations and surrogates to accelerate Monte-Carlo convergence introduced in \carpool, and demonstrate it on multiple $N$-body simulation outputs. Our generalization uses a matrix vectorization procedure (\textit{c.f.,} Figure  \ref{fig:vectorization}). All that is required to CARPool is a surrogate solver that can rapidly generate particle snapshots with minimal computational effort and a surrogate statistic which is strongly correlated with the one computed by the simulation code.  As in \carpool, CARPooling guarantees unbiased results for whatever quantity it is used to estimate, by construction, even if the surrogate is highly biased. No modifications to the simulation codes are required. In this paper, we pair \texttt{GADGET} and \texttt{COLA} to CARPool the elements of the covariance matrix of statistics derived from  \textit{GADGET} $N$-body simulations. As in \carpool\ we do not perform any pre-processing of the statistics to improve correlations to study the raw performance of CARPool. For particular applications there are likely physically motivated approaches that would improve the variance reduction, as discussed, \textit{e.g.}, in section \ref{sec:matterPDF}.

Our presentation focuses on the relative reduction of the Monte-Carlo variance on the covariance matrix, and the resulting reduction in computational cost. An alternative but equivalent perspective is to view CARPool as a technique to de-biases sets of fast surrogates using a limited number of full simulations. Rather than merely accelerating convergence, CARPool can be an enabling technology to help reach accuracy requirements that would have otherwise been out of reach with a fixed amount of computation, since that is likely to be the limited resource. 

We assess the impact of the CARPool variance reduction using derived properties of the covariance matrix, such as the inverse covariance (or precision) matrix, the eigenvalues, a log-likelihood statistic and the (inverse) Fisher matrix. \footnote{The guaranteed lack of bias does not automatically also apply to these transformed quantities. An alternative approach would have been directly to CARPool the derived quantities since the unbiasedness guarantee would then apply. Since the CARPool approach is very flexible, we leave the choice of what quantity to apply it to up to the user.}

For the power spectrum and both equilateral and squeezed triangle configurations of the bispectrum, the variance of the covariance matrix elements is reduced by more than order of magnitude and up to four orders of magnitude. These improvements translate to significantly more accurate log-likelihoods computed from test data, eigenvalues, and estimated confidence regions for cosmological parameters when compared to computations using a reference covariance matrix based on 12,000 $N$-body realisations.

CARPool also gave significant reductions of variance for the covariance matrices of the matter correlation function and the matter one-point probability distribution function (PDF), two real-space clustering statistics. In this instance we found that these improvements did not translate to significant improvements in the derived (test) quantities. For example, in these cases, the eigenvalue ratios and Fisher matrices computed using the CARPool covariance show only marginal improvement compared to those using the sample covariance matrix based on the same number of simulations. These examples have severely ill-conditioned covariance matrices, sometimes causing the smallest eigenvalues of the CARPool estimator to scatter negative. This can be understood because our matrix generalization of CARPool reduces variance on the elements of the covariance matrix but does not explicitly guarantee positive (semi-)definiteness of the covariance matrix. A possible way to address this issue would be to apply techniques designed to improve the condition of the covariance matrix such as tapering  \citep{10.1093/mnras/stv2259}  and shrinkage estimators  \citep{schafer2005,2008MNRAS.389..766P,2017MNRAS.466L..83J}, albeit at the cost of giving up the strict unbiasedness of the covariance matrix estimates. Alternatively, we could seek to impose positive definiteness in our matrix generalization of the CARPool method; this will be subject of future work.

More generally, combining CARPool with other fast estimators of the covariance matrix listed in section \ref{sec:introduction} will likely lead to further improvements since CARPool relies on the entirely different principle of combining simulations and surrogates. Similarly, we explored only a single surrogate. Combinations of surrogates or better surrogates, such as those mentioned in section \ref{sec:introduction} could well lead to further acceleration. We leave an exploration of such combined estimators to future work.
 
\section*{Acknowledgements}
N.C. acknowledges funding from LabEx ENS-ICFP (PSL). B.D.W. acknowledges support by the ANR BIG4 project, grant ANR-16-CE23-0002 of the French Agence Nationale de la Recherche;  and  the Labex ILP (reference ANR-10-LABX-63) part of the Idex SUPER, and received financial state aid managed by the Agence Nationale de la Recherche, as part of the programme Investissements d'avenir under the reference ANR-11-IDEX-0004-02.
The Flatiron Institute is supported by the Simons Foundation.

\section*{Data Availability}

The data underlying this article are available through \textit{globus.org}, and instructions can be found at \url{https://github.com/franciscovillaescusa/Quijote-simulations}. Additionally, a \texttt{Python3} package with code examples and documentation is provided at \url{https://github.com/CompiledAtBirth/pyCARPool} to experiment with CARPool.



\bibliographystyle{mnras}
\bibliography{covBBL}

\begin{thebibliography}{}
\makeatletter
\relax
\def\mn@urlcharsother{\let\do\@makeother \do\$\do\&\do\#\do\^\do\_\do\%\do\~}
\def\mn@doi{\begingroup\mn@urlcharsother \@ifnextchar [ {\mn@doi@}
  {\mn@doi@[]}}
\def\mn@doi@[#1]#2{\def\@tempa{#1}\ifx\@tempa\@empty \href
  {http://dx.doi.org/#2} {doi:#2}\else \href {http://dx.doi.org/#2} {#1}\fi
  \endgroup}
\def\mn@eprint#1#2{\mn@eprint@#1:#2::\@nil}
\def\mn@eprint@arXiv#1{\href {http://arxiv.org/abs/#1} {{\tt arXiv:#1}}}
\def\mn@eprint@dblp#1{\href {http://dblp.uni-trier.de/rec/bibtex/#1.xml}
  {dblp:#1}}
\def\mn@eprint@#1:#2:#3:#4\@nil{\def\@tempa {#1}\def\@tempb {#2}\def\@tempc
  {#3}\ifx \@tempc \@empty \let \@tempc \@tempb \let \@tempb \@tempa \fi \ifx
  \@tempb \@empty \def\@tempb {arXiv}\fi \@ifundefined
  {mn@eprint@\@tempb}{\@tempb:\@tempc}{\expandafter \expandafter \csname
  mn@eprint@\@tempb\endcsname \expandafter{\@tempc}}}

\bibitem[\protect\citeauthoryear{{Alsing} \& {Wandelt}}{{Alsing} \&
  {Wandelt}}{2018}]{alsingwandelt2018}
{Alsing} J.,  {Wandelt} B.,  2018, \mn@doi [\mnras] {10.1093/mnrasl/sly029},
  \href {https://ui.adsabs.harvard.edu/abs/2018MNRAS.476L..60A} {476, L60}

\bibitem[\protect\citeauthoryear{{Angulo} \& {Pontzen}}{{Angulo} \&
  {Pontzen}}{2016}]{2016MNRAS.462L...1A}
{Angulo} R.~E.,  {Pontzen} A.,  2016, \mn@doi [\mnras] {10.1093/mnrasl/slw098},
  \href {https://ui.adsabs.harvard.edu/abs/2016MNRAS.462L...1A} {462, L1}

\bibitem[\protect\citeauthoryear{{Angulo}, {Zennaro}, {Contreras}, {Aric{\`o}},
  {Pellejero-Iba{\~n}ez}  \& {St{\"u}cker}}{{Angulo}
  et~al.}{2020}]{2020arXiv200406245A}
{Angulo} R.~E.,  {Zennaro} M.,  {Contreras} S.,  {Aric{\`o}} G.,
  {Pellejero-Iba{\~n}ez} M.,   {St{\"u}cker} J.,  2020, arXiv e-prints, \href
  {https://ui.adsabs.harvard.edu/abs/2020arXiv200406245A} {p. arXiv:2004.06245}

\bibitem[\protect\citeauthoryear{Bai \& Yin}{Bai \& Yin}{1993}]{baiYin}
Bai Z.~D.,  Yin Y.~Q.,  1993, \mn@doi [The Annals of Probability]
  {10.1214/aop/1176989118}, 21, 1275

\bibitem[\protect\citeauthoryear{Bernardeau, Colombi, Gaztanaga  \&
  Scoccimarro}{Bernardeau et~al.}{2002}]{Bernardeau:2001qr}
Bernardeau F.,  Colombi S.,  Gaztanaga E.,   Scoccimarro R.,  2002, \mn@doi
  [Phys. Rept.] {10.1016/S0370-1573(02)00135-7}, 367, 1

\bibitem[\protect\citeauthoryear{Blot, Corasaniti, Alimi, Reverdy  \&
  Rasera}{Blot et~al.}{2014}]{10.1093/mnras/stu2190}
Blot L.,  Corasaniti P.~S.,  Alimi J.-M.,  Reverdy V.,   Rasera Y.,  2014,
  \mn@doi [Monthly Notices of the Royal Astronomical Society]
  {10.1093/mnras/stu2190}, 446, 1756

\bibitem[\protect\citeauthoryear{{Blot}, {Corasaniti}, {Amendola}  \&
  {Kitching}}{{Blot} et~al.}{2016}]{2016MNRAS.458.4462B}
{Blot} L.,  {Corasaniti} P.~S.,  {Amendola} L.,   {Kitching} T.~D.,  2016,
  \mn@doi [\mnras] {10.1093/mnras/stw604}, \href
  {https://ui.adsabs.harvard.edu/abs/2016MNRAS.458.4462B} {458, 4462}

\bibitem[\protect\citeauthoryear{{Blot} et~al.,}{{Blot}
  et~al.}{2019}]{2019MNRAS.485.2806B}
{Blot} L.,  et~al., 2019, \mn@doi [\mnras] {10.1093/mnras/stz507}, \href
  {https://ui.adsabs.harvard.edu/abs/2019MNRAS.485.2806B} {485, 2806}

\bibitem[\protect\citeauthoryear{{Carron}}{{Carron}}{2013}]{carron2013}
{Carron} J.,  2013, \mn@doi [\aap] {10.1051/0004-6361/201220538}, \href
  {https://ui.adsabs.harvard.edu/abs/2013A&A...551A..88C} {551, A88}

\bibitem[\protect\citeauthoryear{{Chartier}, {Wandelt}, {Akrami}  \&
  {Villaescusa-Navarro}}{{Chartier} et~al.}{2020}]{2020arXiv200908970C}
{Chartier} N.,  {Wandelt} B.,  {Akrami} Y.,   {Villaescusa-Navarro} F.,  2020,
  arXiv e-prints, \href {https://ui.adsabs.harvard.edu/abs/2020arXiv200908970C}
  {p. arXiv:2009.08970}

\bibitem[\protect\citeauthoryear{{Chuang}, {Kitaura}, {Prada}, {Zhao}  \&
  {Yepes}}{{Chuang} et~al.}{2015}]{2015MNRAS.446.2621C}
{Chuang} C.-H.,  {Kitaura} F.-S.,  {Prada} F.,  {Zhao} C.,   {Yepes} G.,  2015,
  \mn@doi [\mnras] {10.1093/mnras/stu2301}, \href
  {https://ui.adsabs.harvard.edu/abs/2015MNRAS.446.2621C} {446, 2621}

\bibitem[\protect\citeauthoryear{{Colavincenzo} et~al.,}{{Colavincenzo}
  et~al.}{2019}]{2019MNRAS.482.4883C}
{Colavincenzo} M.,  et~al., 2019, \mn@doi [\mnras] {10.1093/mnras/sty2964},
  \href {https://ui.adsabs.harvard.edu/abs/2019MNRAS.482.4883C} {482, 4883}

\bibitem[\protect\citeauthoryear{{Dai} \& {Seljak}}{{Dai} \&
  {Seljak}}{2020}]{2020arXiv201002926D}
{Dai} B.,  {Seljak} U.,  2020, arXiv e-prints, \href
  {https://ui.adsabs.harvard.edu/abs/2020arXiv201002926D} {p. arXiv:2010.02926}

\bibitem[\protect\citeauthoryear{DeRose et~al.,}{DeRose
  et~al.}{2019}]{DeRose_2019}
DeRose J.,  et~al., 2019, \mn@doi [The Astrophysical Journal]
  {10.3847/1538-4357/ab1085}, 875, 69

\bibitem[\protect\citeauthoryear{Desjacques, Jeong  \& Schmidt}{Desjacques
  et~al.}{2018}]{Desjacques:2016bnm}
Desjacques V.,  Jeong D.,   Schmidt F.,  2018, \mn@doi [Phys. Rept.]
  {10.1016/j.physrep.2017.12.002}, 733, 1

\bibitem[\protect\citeauthoryear{{Dodelson} \& {Schneider}}{{Dodelson} \&
  {Schneider}}{2013}]{2013PhRvD..88f3537D}
{Dodelson} S.,  {Schneider} M.~D.,  2013, \mn@doi [\prd]
  {10.1103/PhysRevD.88.063537}, \href
  {https://ui.adsabs.harvard.edu/abs/2013PhRvD..88f3537D} {88, 063537}

\bibitem[\protect\citeauthoryear{{Eifler, T.}, {Schneider, P.}  \& {Hartlap,
  J.}}{{Eifler, T.} et~al.}{2009}]{refId0}
{Eifler, T.} {Schneider, P.}  {Hartlap, J.} 2009, \mn@doi [A\&A]
  {10.1051/0004-6361/200811276}, 502, 721

\bibitem[\protect\citeauthoryear{{Escoffier} et~al.,}{{Escoffier}
  et~al.}{2016}]{2016arXiv160600233E}
{Escoffier} S.,  et~al., 2016, arXiv e-prints, \href
  {https://ui.adsabs.harvard.edu/abs/2016arXiv160600233E} {p. arXiv:1606.00233}

\bibitem[\protect\citeauthoryear{{Favole}, {Granett}, {Silva Lafaurie}  \&
  {Sapone}}{{Favole} et~al.}{2020}]{2020arXiv200413436F}
{Favole} G.,  {Granett} B.~R.,  {Silva Lafaurie} J.,   {Sapone} D.,  2020,
  arXiv e-prints, \href {https://ui.adsabs.harvard.edu/abs/2020arXiv200413436F}
  {p. arXiv:2004.13436}

\bibitem[\protect\citeauthoryear{{Feng}, {Chu}, {Seljak}  \& {McDonald}}{{Feng}
  et~al.}{2016}]{2016MNRAS.463.2273F}
{Feng} Y.,  {Chu} M.-Y.,  {Seljak} U.,   {McDonald} P.,  2016, \mn@doi [\mnras]
  {10.1093/mnras/stw2123}, \href
  {https://ui.adsabs.harvard.edu/abs/2016MNRAS.463.2273F} {463, 2273}

\bibitem[\protect\citeauthoryear{{Friedrich} \& {Eifler}}{{Friedrich} \&
  {Eifler}}{2018}]{2018MNRAS.473.4150F}
{Friedrich} O.,  {Eifler} T.,  2018, \mn@doi [\mnras] {10.1093/mnras/stx2566},
  \href {https://ui.adsabs.harvard.edu/abs/2018MNRAS.473.4150F} {473, 4150}

\bibitem[\protect\citeauthoryear{{Friedrich} et~al.,}{{Friedrich}
  et~al.}{2021}]{2021MNRAS.tmp.2208F}
{Friedrich} O.,  et~al., 2021, \mn@doi [\mnras] {10.1093/mnras/stab2384}, \href
  {https://ui.adsabs.harvard.edu/abs/2021MNRAS.tmp.2208F} {}

\bibitem[\protect\citeauthoryear{Garrison}{Garrison}{2019}]{phdAbacus}
Garrison L.,  2019, PhD thesis, University Of Washington

\bibitem[\protect\citeauthoryear{Glynn \& Szechtman}{Glynn \&
  Szechtman}{2002}]{10.1007/978-3-642-56046-0_3}
Glynn P.~W.,  Szechtman R.,  2002, in Fang K.-T.,  Niederreiter H.,
  Hickernell F.~J.,  eds, Monte Carlo and Quasi-Monte Carlo Methods 2000.
  Springer Berlin Heidelberg, Berlin, Heidelberg, pp 27--49

\bibitem[\protect\citeauthoryear{{Habib} et~al.,}{{Habib}
  et~al.}{2016}]{2016NewA...42...49H}
{Habib} S.,  et~al., 2016, \mn@doi [\na] {10.1016/j.newast.2015.06.003}, \href
  {https://ui.adsabs.harvard.edu/abs/2016NewA...42...49H} {42, 49}

\bibitem[\protect\citeauthoryear{{Hall} \& {Taylor}}{{Hall} \&
  {Taylor}}{2019}]{2019MNRAS.483..189H}
{Hall} A.,  {Taylor} A.,  2019, \mn@doi [\mnras] {10.1093/mnras/sty3102}, \href
  {https://ui.adsabs.harvard.edu/abs/2019MNRAS.483..189H} {483, 189}

\bibitem[\protect\citeauthoryear{{Harnois-D{\'e}raps} \&
  {Pen}}{{Harnois-D{\'e}raps} \& {Pen}}{2013}]{2013MNRAS.431.3349H}
{Harnois-D{\'e}raps} J.,  {Pen} U.-L.,  2013, \mn@doi [\mnras]
  {10.1093/mnras/stt413}, \href
  {https://ui.adsabs.harvard.edu/abs/2013MNRAS.431.3349H} {431, 3349}

\bibitem[\protect\citeauthoryear{{Harnois-D{\'e}raps}, {Vafaei}  \& {Van
  Waerbeke}}{{Harnois-D{\'e}raps} et~al.}{2012}]{2012MNRAS.426.1262H}
{Harnois-D{\'e}raps} J.,  {Vafaei} S.,   {Van Waerbeke} L.,  2012, \mn@doi
  [\mnras] {10.1111/j.1365-2966.2012.21624.x}, \href
  {https://ui.adsabs.harvard.edu/abs/2012MNRAS.426.1262H} {426, 1262}

\bibitem[\protect\citeauthoryear{{Harnois-D{\'e}raps}, {Giblin}  \&
  {Joachimi}}{{Harnois-D{\'e}raps} et~al.}{2019}]{2019A&A...631A.160H}
{Harnois-D{\'e}raps} J.,  {Giblin} B.,   {Joachimi} B.,  2019, \mn@doi [\aap]
  {10.1051/0004-6361/201935912}, \href
  {https://ui.adsabs.harvard.edu/abs/2019A&A...631A.160H} {631, A160}

\bibitem[\protect\citeauthoryear{{Hartlap}, {Simon}  \& {Schneider}}{{Hartlap}
  et~al.}{2007}]{2007A&A...464..399H}
{Hartlap} J.,  {Simon} P.,   {Schneider} P.,  2007, \mn@doi [\aap]
  {10.1051/0004-6361:20066170}, \href
  {https://ui.adsabs.harvard.edu/abs/2007A&A...464..399H} {464, 399}

\bibitem[\protect\citeauthoryear{He, Li, Feng, Ho, Ravanbakhsh, Chen  \&
  Póczos}{He et~al.}{2019}]{He_2019}
He S.,  Li Y.,  Feng Y.,  Ho S.,  Ravanbakhsh S.,  Chen W.,   Póczos B.,
  2019, \mn@doi [Proceedings of the National Academy of Sciences]
  {10.1073/pnas.1821458116}, 116, 13825–13832

\bibitem[\protect\citeauthoryear{{Heavens}, {Jimenez}  \& {Lahav}}{{Heavens}
  et~al.}{2000}]{moped}
{Heavens} A.~F.,  {Jimenez} R.,   {Lahav} O.,  2000, \mn@doi [\mnras]
  {10.1046/j.1365-8711.2000.03692.x}, \href
  {https://ui.adsabs.harvard.edu/abs/2000MNRAS.317..965H} {317, 965}

\bibitem[\protect\citeauthoryear{{Hikage}, {Takahashi}  \& {Koyama}}{{Hikage}
  et~al.}{2020}]{2020PhRvD.102h3514H}
{Hikage} C.,  {Takahashi} R.,   {Koyama} K.,  2020, \mn@doi [\prd]
  {10.1103/PhysRevD.102.083514}, \href
  {https://ui.adsabs.harvard.edu/abs/2020PhRvD.102h3514H} {102, 083514}

\bibitem[\protect\citeauthoryear{Howlett, Manera  \& Percival}{Howlett
  et~al.}{2015}]{Howlett_2015}
Howlett C.,  Manera M.,   Percival W.,  2015, \mn@doi [Astronomy and Computing]
  {10.1016/j.ascom.2015.07.003}, 12, 109–126

\bibitem[\protect\citeauthoryear{{Ishiyama}, {Fukushige}  \&
  {Makino}}{{Ishiyama} et~al.}{2009}]{2009PASJ...61.1319I}
{Ishiyama} T.,  {Fukushige} T.,   {Makino} J.,  2009, \mn@doi [\pasj]
  {10.1093/pasj/61.6.1319}, \href
  {https://ui.adsabs.harvard.edu/abs/2009PASJ...61.1319I} {61, 1319}

\bibitem[\protect\citeauthoryear{Izard, Crocce  \& Fosalba}{Izard
  et~al.}{2016}]{Izard_2016}
Izard A.,  Crocce M.,   Fosalba P.,  2016, \mn@doi [Monthly Notices of the
  Royal Astronomical Society] {10.1093/mnras/stw797}, 459, 2327–2341

\bibitem[\protect\citeauthoryear{{Joachimi}}{{Joachimi}}{2017}]{2017MNRAS.466L..83J}
{Joachimi} B.,  2017, \mn@doi [\mnras] {10.1093/mnrasl/slw240}, \href
  {https://ui.adsabs.harvard.edu/abs/2017MNRAS.466L..83J} {466, L83}

\bibitem[\protect\citeauthoryear{{Kasim} et~al.,}{{Kasim}
  et~al.}{2020}]{2020arXiv200108055K}
{Kasim} M.~F.,  et~al., 2020, arXiv e-prints, \href
  {https://ui.adsabs.harvard.edu/abs/2020arXiv200108055K} {p. arXiv:2001.08055}

\bibitem[\protect\citeauthoryear{{Kitaura}, {Yepes}  \& {Prada}}{{Kitaura}
  et~al.}{2014}]{2014MNRAS.439L..21K}
{Kitaura} F.~S.,  {Yepes} G.,   {Prada} F.,  2014, \mn@doi [\mnras]
  {10.1093/mnrasl/slt172}, \href
  {https://ui.adsabs.harvard.edu/abs/2014MNRAS.439L..21K} {439, L21}

\bibitem[\protect\citeauthoryear{Kodi~Ramanah, Charnock, Villaescusa-Navarro
  \& Wandelt}{Kodi~Ramanah et~al.}{2020}]{Kodi_Ramanah_2020}
Kodi~Ramanah D.,  Charnock T.,  Villaescusa-Navarro F.,   Wandelt B.~D.,  2020,
  \mn@doi [Monthly Notices of the Royal Astronomical Society]
  {10.1093/mnras/staa1428}, 495, 4227–4236

\bibitem[\protect\citeauthoryear{{Kodwani}, {Alonso}  \& {Ferreira}}{{Kodwani}
  et~al.}{2019}]{kodwani2019}
{Kodwani} D.,  {Alonso} D.,   {Ferreira} P.,  2019, \mn@doi [The Open Journal
  of Astrophysics] {10.21105/astro.1811.11584}, \href
  {https://ui.adsabs.harvard.edu/abs/2019OJAp....2E...3K} {2, 3}

\bibitem[\protect\citeauthoryear{Lavenberg \& Welch}{Lavenberg \&
  Welch}{1981}]{Lavenberg1981APO}
Lavenberg S.,  Welch P.,  1981, Management Science, 27, 322

\bibitem[\protect\citeauthoryear{{Leclercq}, {Jasche}, {Gil-Mar{\'\i}n}  \&
  {Wandelt}}{{Leclercq} et~al.}{2013}]{2013JCAP...11..048L}
{Leclercq} F.,  {Jasche} J.,  {Gil-Mar{\'\i}n} H.,   {Wandelt} B.,  2013,
  \mn@doi [\jcap] {10.1088/1475-7516/2013/11/048}, \href
  {https://ui.adsabs.harvard.edu/abs/2013JCAP...11..048L} {2013, 048}

\bibitem[\protect\citeauthoryear{{Leclercq}, {Faure}, {Lavaux}, {Wandelt},
  {Jaffe}, {Heavens}  \& {Percival}}{{Leclercq}
  et~al.}{2020}]{2020A&A...639A..91L}
{Leclercq} F.,  {Faure} B.,  {Lavaux} G.,  {Wandelt} B.~D.,  {Jaffe} A.~H.,
  {Heavens} A.~F.,   {Percival} W.~J.,  2020, \mn@doi [\aap]
  {10.1051/0004-6361/202037995}, \href
  {https://ui.adsabs.harvard.edu/abs/2020A&A...639A..91L} {639, A91}

\bibitem[\protect\citeauthoryear{{Li}, {Singh}, {Yu}, {Feng}  \& {Seljak}}{{Li}
  et~al.}{2019}]{2019JCAP...01..016L}
{Li} Y.,  {Singh} S.,  {Yu} B.,  {Feng} Y.,   {Seljak} U.,  2019, \mn@doi
  [\jcap] {10.1088/1475-7516/2019/01/016}, \href
  {https://ui.adsabs.harvard.edu/abs/2019JCAP...01..016L} {2019, 016}

\bibitem[\protect\citeauthoryear{{Lippich} et~al.,}{{Lippich}
  et~al.}{2019}]{2019MNRAS.482.1786L}
{Lippich} M.,  et~al., 2019, \mn@doi [\mnras] {10.1093/mnras/sty2757}, \href
  {https://ui.adsabs.harvard.edu/abs/2019MNRAS.482.1786L} {482, 1786}

\bibitem[\protect\citeauthoryear{{McClintock} et~al.,}{{McClintock}
  et~al.}{2019a}]{2019arXiv190713167M}
{McClintock} T.,  et~al., 2019a, arXiv e-prints, \href
  {https://ui.adsabs.harvard.edu/abs/2019arXiv190713167M} {p. arXiv:1907.13167}

\bibitem[\protect\citeauthoryear{McClintock et~al.,}{McClintock
  et~al.}{2019b}]{McClintock_2019}
McClintock T.,  et~al., 2019b, \mn@doi [The Astrophysical Journal]
  {10.3847/1538-4357/aaf568}, 872, 53

\bibitem[\protect\citeauthoryear{{Modi}, {Lanusse}  \& {Seljak}}{{Modi}
  et~al.}{2020}]{2020arXiv201011847M}
{Modi} C.,  {Lanusse} F.,   {Seljak} U.,  2020, arXiv e-prints, \href
  {https://ui.adsabs.harvard.edu/abs/2020arXiv201011847M} {p. arXiv:2010.11847}

\bibitem[\protect\citeauthoryear{Mohammed \& Seljak}{Mohammed \&
  Seljak}{2014}]{Mohammed:2014lja}
Mohammed I.,  Seljak U.,  2014, \mn@doi [Mon. Not. Roy. Astron. Soc.]
  {10.1093/mnras/stu1972}, 445, 3382

\bibitem[\protect\citeauthoryear{{Mohammed}, {Seljak}  \& {Vlah}}{{Mohammed}
  et~al.}{2017}]{2017MNRAS.466..780M}
{Mohammed} I.,  {Seljak} U.,   {Vlah} Z.,  2017, \mn@doi [\mnras]
  {10.1093/mnras/stw3196}, \href
  {https://ui.adsabs.harvard.edu/abs/2017MNRAS.466..780M} {466, 780}

\bibitem[\protect\citeauthoryear{{Monaco}, {Sefusatti}, {Borgani}, {Crocce},
  {Fosalba}, {Sheth}  \& {Theuns}}{{Monaco} et~al.}{2013}]{2013MNRAS.433.2389M}
{Monaco} P.,  {Sefusatti} E.,  {Borgani} S.,  {Crocce} M.,  {Fosalba} P.,
  {Sheth} R.~K.,   {Theuns} T.,  2013, \mn@doi [\mnras] {10.1093/mnras/stt907},
  \href {https://ui.adsabs.harvard.edu/abs/2013MNRAS.433.2389M} {433, 2389}

\bibitem[\protect\citeauthoryear{Paz \& Sánchez}{Paz \&
  Sánchez}{2015}]{10.1093/mnras/stv2259}
Paz D.~J.,  Sánchez A.~G.,  2015, \mn@doi [Monthly Notices of the Royal
  Astronomical Society] {10.1093/mnras/stv2259}, 454, 4326

\bibitem[\protect\citeauthoryear{{Pearson} \& {Samushia}}{{Pearson} \&
  {Samushia}}{2016}]{2016MNRAS.457..993P}
{Pearson} D.~W.,  {Samushia} L.,  2016, \mn@doi [\mnras]
  {10.1093/mnras/stw062}, \href
  {https://ui.adsabs.harvard.edu/abs/2016MNRAS.457..993P} {457, 993}

\bibitem[\protect\citeauthoryear{{Percival} et~al.,}{{Percival}
  et~al.}{2014}]{2014MNRAS.439.2531P}
{Percival} W.~J.,  et~al., 2014, \mn@doi [\mnras] {10.1093/mnras/stu112}, \href
  {https://ui.adsabs.harvard.edu/abs/2014MNRAS.439.2531P} {439, 2531}

\bibitem[\protect\citeauthoryear{{Percival}, {Friedrich}, {Sellentin}  \&
  {Heavens}}{{Percival} et~al.}{2021}]{2021arXiv210810402P}
{Percival} W.~J.,  {Friedrich} O.,  {Sellentin} E.,   {Heavens} A.,  2021,
  arXiv e-prints, \href {https://ui.adsabs.harvard.edu/abs/2021arXiv210810402P}
  {p. arXiv:2108.10402}

\bibitem[\protect\citeauthoryear{{Philcox} \& {Eisenstein}}{{Philcox} \&
  {Eisenstein}}{2019}]{2019MNRAS.490.5931P}
{Philcox} O. H.~E.,  {Eisenstein} D.~J.,  2019, \mn@doi [\mnras]
  {10.1093/mnras/stz2896}, \href
  {https://ui.adsabs.harvard.edu/abs/2019MNRAS.490.5931P} {490, 5931}

\bibitem[\protect\citeauthoryear{{Philcox}, {Eisenstein}, {O'Connell}  \&
  {Wiegand}}{{Philcox} et~al.}{2020}]{2020MNRAS.491.3290P}
{Philcox} O. H.~E.,  {Eisenstein} D.~J.,  {O'Connell} R.,   {Wiegand} A.,
  2020, \mn@doi [\mnras] {10.1093/mnras/stz3218}, \href
  {https://ui.adsabs.harvard.edu/abs/2020MNRAS.491.3290P} {491, 3290}

\bibitem[\protect\citeauthoryear{{Philcox}, {Ivanov}, {Zaldarriaga},
  {Simonovi{\'c}}  \& {Schmittfull}}{{Philcox}
  et~al.}{2021}]{2021PhRvD.103d3508P}
{Philcox} O. H.~E.,  {Ivanov} M.~M.,  {Zaldarriaga} M.,  {Simonovi{\'c}} M.,
  {Schmittfull} M.,  2021, \mn@doi [\prd] {10.1103/PhysRevD.103.043508}, \href
  {https://ui.adsabs.harvard.edu/abs/2021PhRvD.103d3508P} {103, 043508}

\bibitem[\protect\citeauthoryear{{Planck Collaboration} et~al.,}{{Planck
  Collaboration} et~al.}{2020}]{2020A&A...641A...6P}
{Planck Collaboration} et~al., 2020, \mn@doi [\aap]
  {10.1051/0004-6361/201833910}, \href
  {https://ui.adsabs.harvard.edu/abs/2020A&A...641A...6P} {641, A6}

\bibitem[\protect\citeauthoryear{{Pontzen}, {Slosar}, {Roth}  \&
  {Peiris}}{{Pontzen} et~al.}{2016}]{2016PhRvD..93j3519P}
{Pontzen} A.,  {Slosar} A.,  {Roth} N.,   {Peiris} H.~V.,  2016, \mn@doi [\prd]
  {10.1103/PhysRevD.93.103519}, \href
  {https://ui.adsabs.harvard.edu/abs/2016PhRvD..93j3519P} {93, 103519}

\bibitem[\protect\citeauthoryear{{Pope} \& {Szapudi}}{{Pope} \&
  {Szapudi}}{2008}]{2008MNRAS.389..766P}
{Pope} A.~C.,  {Szapudi} I.,  2008, \mn@doi [\mnras]
  {10.1111/j.1365-2966.2008.13561.x}, \href
  {https://ui.adsabs.harvard.edu/abs/2008MNRAS.389..766P} {389, 766}

\bibitem[\protect\citeauthoryear{{Potter}, {Stadel}  \& {Teyssier}}{{Potter}
  et~al.}{2017}]{2017ComAC...4....2P}
{Potter} D.,  {Stadel} J.,   {Teyssier} R.,  2017, \mn@doi [Computational
  Astrophysics and Cosmology] {10.1186/s40668-017-0021-1}, \href
  {https://ui.adsabs.harvard.edu/abs/2017ComAC...4....2P} {4, 2}

\bibitem[\protect\citeauthoryear{Rubinstein \& Marcus}{Rubinstein \&
  Marcus}{1985}]{10.1287/opre.33.3.661}
Rubinstein R.~Y.,  Marcus R.,  1985, \mn@doi [Oper. Res.]
  {10.1287/opre.33.3.661}, 33, 661–677

\bibitem[\protect\citeauthoryear{Schäfer \& Strimmer}{Schäfer \&
  Strimmer}{2005}]{schafer2005}
Schäfer J.,  Strimmer K.,  2005, \mn@doi [Statistical applications in genetics
  and molecular biology] {10.2202/1544-6115.1175}, 4, Article32

\bibitem[\protect\citeauthoryear{{Scoccimarro} \& {Sheth}}{{Scoccimarro} \&
  {Sheth}}{2002}]{2002MNRAS.329..629S}
{Scoccimarro} R.,  {Sheth} R.~K.,  2002, \mn@doi [\mnras]
  {10.1046/j.1365-8711.2002.04999.x}, \href
  {https://ui.adsabs.harvard.edu/abs/2002MNRAS.329..629S} {329, 629}

\bibitem[\protect\citeauthoryear{{Sellentin} \& {Heavens}}{{Sellentin} \&
  {Heavens}}{2018}]{2018MNRAS.473.2355S}
{Sellentin} E.,  {Heavens} A.~F.,  2018, \mn@doi [\mnras]
  {10.1093/mnras/stx2491}, \href
  {https://ui.adsabs.harvard.edu/abs/2018MNRAS.473.2355S} {473, 2355}

\bibitem[\protect\citeauthoryear{{Springel}}{{Springel}}{2005}]{2005MNRAS.364.1105S}
{Springel} V.,  2005, \mn@doi [\mnras] {10.1111/j.1365-2966.2005.09655.x},
  \href {https://ui.adsabs.harvard.edu/abs/2005MNRAS.364.1105S} {364, 1105}

\bibitem[\protect\citeauthoryear{Taffoni, Monaco  \& Theuns}{Taffoni
  et~al.}{2002}]{pino}
Taffoni G.,  Monaco P.,   Theuns T.,  2002, \mn@doi [Monthly Notices of the
  Royal Astronomical Society] {10.1046/j.1365-8711.2002.05441.x}, 333, 623

\bibitem[\protect\citeauthoryear{{Takahashi} et~al.,}{{Takahashi}
  et~al.}{2009}]{2009ApJ...700..479T}
{Takahashi} R.,  et~al., 2009, \mn@doi [\apj] {10.1088/0004-637X/700/1/479},
  \href {https://ui.adsabs.harvard.edu/abs/2009ApJ...700..479T} {700, 479}

\bibitem[\protect\citeauthoryear{{Tassev} \& {Zaldarriaga}}{{Tassev} \&
  {Zaldarriaga}}{2012}]{2012JCAP...04..013T}
{Tassev} S.,  {Zaldarriaga} M.,  2012, \mn@doi [\jcap]
  {10.1088/1475-7516/2012/04/013}, \href
  {https://ui.adsabs.harvard.edu/abs/2012JCAP...04..013T} {2012, 013}

\bibitem[\protect\citeauthoryear{Tassev, Zaldarriaga  \& Eisenstein}{Tassev
  et~al.}{2013}]{Tassev_2013}
Tassev S.,  Zaldarriaga M.,   Eisenstein D.~J.,  2013, \mn@doi [Journal of
  Cosmology and Astroparticle Physics] {10.1088/1475-7516/2013/06/036}, 2013,
  036–036

\bibitem[\protect\citeauthoryear{{Tassev}, {Eisenstein}, {Wand elt}  \&
  {Zaldarriaga}}{{Tassev} et~al.}{2015}]{2015arXiv150207751T}
{Tassev} S.,  {Eisenstein} D.~J.,  {Wand elt} B.~D.,   {Zaldarriaga} M.,  2015,
  arXiv e-prints, \href {https://ui.adsabs.harvard.edu/abs/2015arXiv150207751T}
  {p. arXiv:1502.07751}

\bibitem[\protect\citeauthoryear{{Taylor} \& {Joachimi}}{{Taylor} \&
  {Joachimi}}{2014}]{2014MNRAS.442.2728T}
{Taylor} A.,  {Joachimi} B.,  2014, \mn@doi [\mnras] {10.1093/mnras/stu996},
  \href {https://ui.adsabs.harvard.edu/abs/2014MNRAS.442.2728T} {442, 2728}

\bibitem[\protect\citeauthoryear{{Taylor}, {Joachimi}  \& {Kitching}}{{Taylor}
  et~al.}{2013}]{2013MNRAS.432.1928T}
{Taylor} A.,  {Joachimi} B.,   {Kitching} T.,  2013, \mn@doi [\mnras]
  {10.1093/mnras/stt270}, \href
  {https://ui.adsabs.harvard.edu/abs/2013MNRAS.432.1928T} {432, 1928}

\bibitem[\protect\citeauthoryear{Villaescusa-Navarro
  et~al.,}{Villaescusa-Navarro et~al.}{2018}]{Villaescusa_Navarro_2018}
Villaescusa-Navarro F.,  et~al., 2018, \mn@doi [The Astrophysical Journal]
  {10.3847/1538-4357/aae52b}, 867, 137

\bibitem[\protect\citeauthoryear{{Villaescusa-Navarro}
  et~al.,}{{Villaescusa-Navarro} et~al.}{2020}]{2020ApJS..250....2V}
{Villaescusa-Navarro} F.,  et~al., 2020, \mn@doi [\apjs]
  {10.3847/1538-4365/ab9d82}, \href
  {https://ui.adsabs.harvard.edu/abs/2020ApJS..250....2V} {250, 2}

\bibitem[\protect\citeauthoryear{{Wadekar}, {Ivanov}  \&
  {Scoccimarro}}{{Wadekar} et~al.}{2020}]{2020PhRvD.102l3521W}
{Wadekar} D.,  {Ivanov} M.~M.,   {Scoccimarro} R.,  2020, \mn@doi [\prd]
  {10.1103/PhysRevD.102.123521}, \href
  {https://ui.adsabs.harvard.edu/abs/2020PhRvD.102l3521W} {102, 123521}

\bibitem[\protect\citeauthoryear{{Warren}}{{Warren}}{2013}]{2013arXiv1310.4502W}
{Warren} M.~S.,  2013, arXiv e-prints, \href
  {https://ui.adsabs.harvard.edu/abs/2013arXiv1310.4502W} {p. arXiv:1310.4502}

\bibitem[\protect\citeauthoryear{{White}, {Tinker}  \& {McBride}}{{White}
  et~al.}{2014}]{2014MNRAS.437.2594W}
{White} M.,  {Tinker} J.~L.,   {McBride} C.~K.,  2014, \mn@doi [\mnras]
  {10.1093/mnras/stt2071}, \href
  {https://ui.adsabs.harvard.edu/abs/2014MNRAS.437.2594W} {437, 2594}

\bibitem[\protect\citeauthoryear{Zhai et~al.,}{Zhai et~al.}{2019}]{Zhai_2019}
Zhai Z.,  et~al., 2019, \mn@doi [The Astrophysical Journal]
  {10.3847/1538-4357/ab0d7b}, 874, 95

\makeatother
\end{thebibliography}








\bsp	
\label{lastpage}
\end{document}